\documentclass[fleqn,10pt]{wlscirep}
\usepackage[utf8]{inputenc}
\usepackage[T1]{fontenc}
\usepackage{multirow}
\usepackage{ulem}
\usepackage{soul}
\begin{document}
\title{Ion-acoustic solitons   in  a relativistic Fermi plasma  at finite temperature}

\author[1,a]{Rupak Dey}
\author[2,b]{Gadadhar Banerjee}
\author[1,*]{Amar Prasad Misra}
\author[3,c]{Chandan Bhowmik}
\affil[1]{Department of Mathematics, Siksha Bhavana, Visva-Bharati University, Santiniketan-731 235,  India}
%\affil[2]{Department of Mathematics, University of Engineering \& Management (UEM), Kolkata-700 160, India }
\affil[2]{Department of Mathematics, Burdwan Raj College, University of Burdwan, Burdwan-713 104, India}
\affil[3]{Department of Mathematics, Saheed Nurul Islam Mahavidyalaya, Tentulia-743 286, India}
\affil[*]{Corresponding author; Email: apmisra@visva-bharati.ac.in }
\affil[a]{Email: rupakdey456@gmail.com}
\affil[b]{Email: gban.iitkgp@gmail.com}
\affil[c]{Email: cbhowmik393@gmail.com}
%%%%%%%%%%%%%%%%
\begin{abstract}
The theory of ion-acoustic solitons in nonrelativistic fully degenerate plasmas and nonrelativistic and ultra-relativistic degenerate plasmas at low temperatures is known. We consider a multi-component relativistic degenerate electron-positron-ion plasma at finite temperatures. Specifically, we focus on the intermediate region where the particle's  thermal energy $(k_BT)$ and the rest mass energy $(mc^2)$ do not differ significantly, i.e., $k_BT\sim mc^2$. However, the Fermi energy $(k_BT_F)$ is larger than the thermal energy and the normalized chemical energy ($\xi=\mu/k_BT$) is positive and finite. Two different parameter regimes with $\beta\equiv k_BT/mc^2<1$ and $\beta>1$, relevant for astrophysical plasmas, are defined, and the existence of small amplitude ion-acoustic solitons in these regimes are studied, including the critical cases where the known KdV (Korteweg-de Vries) theory fails. We show that while the solitons with both the positive (compressive) and negative (rarefactive) potentials coexist in the case of $\beta<1$, only compressive solitons can exist in the other regime $(\beta>1)$. Furthermore, while the rarefactive solitons within the parameter domains of $\beta$  and $\xi$ can evolve with increasing amplitude and hence increasing energy,    the energy of compressive solitons reaches a steady state.  
  
\end{abstract}
\flushbottom
\maketitle
\thispagestyle{empty}

\section{Introduction}
Nonlinear propagation of solitary waves in  electron-positron-ion (e-p-i) plasmas has received significant  research interests for understanding the electrostatic as well as electromagnetic disturbances in various plasma environments \cite{shah2010effect, ur2015electrostatic,atteya2020ion,popel1995ion,banerjee2020large}. The dynamics of collective processes in degenerate dense e-p-i plasmas, which are frequently present in laser-solid interaction experiments \cite{sorbo2018,hurricane2014}, as well as in dense astrophysical objects, such as those in  active galactic nuclei  \cite{fabian2012observational}, white dwarfs  \cite{boshkayev2016equilibrium}, pulsar magnetosphere  \cite{michel1982theory}, the early universe,  neutron star  \cite{de2020highly}, quasars, accretion discs and sun atmosphere  \cite{landau2013statistical}, modifies the existing features of nonlinear waves. In degenerate plasmas, physical parameters including the density, magnetic field and the particle temperature can play significant roles in the evolution of electrostatic and electromagnetic waves \cite{fortov2009extreme}. 
Depending on whether the Fermi energy is much larger than, close to or much smaller than the rest mass energy, degenerate species like electrons and positrons in plasmas may be nonrelativistic, relativistic, or ultra-relativistic. Thus,   it is   desirable to have a  pressure law and the particle distribution that can efficiently and accurately  represent  the relevant physics of relativistic degenerate plasmas. In this context,  the characteristics of linear and nonlinear electrostatic waves in relativistic degenerate e-p-i plasmas have been the focus of various studies over the last few years
 \cite{mamun2010solitary,masood2011electrostatic,ur2015electrostatic,akbari2011propagation}. 
\par 
In degenerate e-p-i plasmas,  as the electrons and positrons are extremely dense, their inter particle distance is comparable to the corresponding thermal de-Broglie wavelength and so they obey 
the Fermi Dirac (FD) distribution instead of the Maxwell-Boltzmann distribution. The FD distribution is mostly used when the thermodynamic temperature $T_j$ of $j$-th species particle is comparable to  the corresponding  Fermi temperature $T_{Fj}$ ($j=e$ for electrons  and $j=p$ for positrons). In particular, the limiting conditions $T_j \gg T_{Fj}$ and $T_j \ll T_{Fj}$, respectively,  correspond to the  non-degenerate  and completely degenerate states. However, in most  real situations,   either $T_j<T_{Fj}$ or $T_j>T_{Fj}$ and there is no as such strict upper limits of the energy levels. Accordingly, the electrons and positrons are  said to be degenerate at finite temperature  or  arbitrarily degenerate or   partially degenerate. The term ``partially" is sometimes used to distinguish with the complete degeneracy. Several authors have studied the linear and nonlinear properties of ion-acoustic waves in nonrelativistic plasmas with arbitrary degeneracy of electrons and/or positrons (See, e.g., Refs. \cite{haas2016,haas2022,dey2022}) and in ultra-relativistic degenerate electron-positron-ion plasmas \cite{rasheed2011}. In other contexts, El-Taibany \textit{et al.} \cite{el-taibany2022} studied the theory of ion-acoustic solitary waves in magnetized quantum electron-positron-ion plasmas using the fluid theory approach. Also, discussed are the linear and nonlinear features of invariant ion-acoustic waves in astrophysical and space plasmas using the kinetic theory approach \cite{saberian2022}. Furthermore, the theory of ion-acoustic waves has been developed to take into account the trapped and Landau quantized electrons through the description of Zakharov–Kuznetsov (ZK) equation \cite{jahangir2021}.  
 
%%%%%%%%%%%%%%%%%%%%%%%%%%%
\par 
The primary elements of white dwarf stars are carbon, oxygen, and completely ionized helium and the typical   particle number density is roughly of the order of  $10^{32}$ $\textit{m}^{-3}$ or more. For these kinds of extremely dense astrophysical objects  where relativistic temperatures are common and particles' velocities approach those of light, the relativistic effects are crucial. Thus, the nonlinear effects in such relativistic  degenerate plasmas at finite temperature are able to provide interesting new insights of localization of electrostatic waves. To the best of our knowledge, no effort has been made to study the characteristics of ion-acoustic solitons in   relativistic   degenerate e-p-i plasmas at finite temperature, especially in the intermediate regimes where the particle's Fermi energy does not significantly differ from the particle's thermal energy and the rest mass energy and the chemical energy is positive and larger than the thermal energy.
We ought to mention that the mechanism for the excitation of ion-acoustic solitons, to be discussed in the manuscript, is not related to the quantum electrodynamics in which the Schwinger limit may be applicable.  Also, this study does not consider the mechanism of electron-positron pair creation and annihilation in a strong electromagnetic field  that may be of the order of the Schwinger field strength or beyond. In astrophysical environments, e.g., inside a white dwarf, the mean energies of electrons and positrons increase  as the stellar mass increases so that for a sufficiently massive white dwarf, the electrons and positrons can be relativistic.  
\par
It is to be noted that although high-density electron-positron pairs are efficiently produced in laser-matter interactions \cite{sorbo2018,hurricane2014} or observed in astrophysical environments \cite{thomas2020}, there are also issues with the electron-positron annihilation rate compared to the time scale of oscillations, especially when plasmas are in local thermodynamic and chemical equilibrium. It has been  shown  that such annihilation rate can  significantly drop  with decreasing values of  the particle temperature $T_j$ and freezes  out at $T_j=16~\rm{KeV}\approx1.85\times10^8$ K \cite{thomas2020}. In the environments of white dwarfs with central temperature $T_j\sim10^7-10^{10}~\rm{K}$   and particle number density $\sim10^{28}-10^{35}~\rm{cm}^{-3}$, the electron-positron annihilation time   can remain  longer than the ion plasma period.  So, for the excitation of ion-acoustic waves in such high density plasmas,   the electron-positron annihilation can be safely neglected. 
\par
There are two basic approaches used to investigate the nonlinear evolution of 
electrostatic  solitary waves: the reductive perturbation  technique  \cite{washimi1966propagation} and the Sagdeev pseudopotential approach  \cite{sagdeev1966cooperative,popel1995ion,banerjee2015pseudopotential}. In the former, the Korteweg-de Vries (KdV) equation is derived to describe  the evolution of one-dimensional solitons.  However, the KdV equation may not be valid when the nonlinear coefficient, say $A$ vanishes or tends to vanish. In these critical situations, the higher-order nonlinear corrections are considered to derive the modified KdV (mKdV) equation (in the case of  $A=0$) and/or the Gardner equation  (in the case of $A\simeq0$) \cite{baumjohann1997advanced}.  Numerous investigations have been made  to study the nonlinear propagation of mKdV and Gardner solitons in multi-component plasmas  with Maxwellian/non-Maxwellian particle distributions   \cite{pakzad2011ion, rehman2016ion,masood2017study, jahangir2022interaction,pradhan2022nonlinear}.
\par
In this work, our aim is to study the existence of ion-acoustic solitons in an intermediate regime of relativistic degenerate plasmas at finite temperature where the Fermi energies of electrons and positrons do not  significantly differ from their thermal energies and the  rest mass energy.  The manuscript is organized in the following way. Section \ref{sec-model} presents the modeling of relativistic degenerate e-p-i plasmas at finite temperatures. It demonstrates the basic set of  fluid equations for the  nonrelativistic classical thermal ions and relativistic degenerate electrons and positrons at finite temperatures. Using the Fermi-Dirac distribution, the number densities of electrons and positrons are also derived in two particular cases, namely $k_BT_j<mc^2$ and      $k_BT_j>mc^2$. The physical regimes for the validity of the model are given in Sec. \ref{sec-phys-regime}. While the linear analysis of ion-acoustic waves is presented in Sec. \ref{sec-linear}, the nonlinear analysis for the evolution of ion-acoustic solitons   is given in Sec. \ref{sec-nonlinear} in the two particular cases ($k_BT_j<mc^2$ and    $k_BT_j>mc^2$). This section has some subsections which discuss the properties of KdV, mKdV and Gardner solitons.  Finally, Sec. \ref{sec-summary-conclu} is left to summarize the  main results and conclude. 
\section{The model} 
\label{sec-model}
%%%%%%%%%%%%%%%%%%%%%%%%%%%%%%%%%
We consider the nonlinear excitation of electrostatic waves at ionic time scale in an unmagnetized  multi-component plasma with relativistic flow of degenerate electrons and positrons at finite temperature and  nonrelativistic classical singly charged positive thermal ions. The dynamics of relativistic electrons and positron fluids in one-dimensional geometry is given by   \cite{misra2018,lee2007,gratton1997}
\begin{equation}
 %\begin{split}
 \frac{\gamma_jH_j}{c^2}\frac{d}{dt} (\gamma_j { v}_j)=-q_jn_j\frac{\partial \phi}{\partial x}-\left(\frac{\partial}{\partial x}+\frac{\gamma_j^2{ v}_j}{c^2}\frac{d}{d t}\right)P_j, \label{eq1}
 %\end{split}
 \end{equation}
 \begin{equation}
 \frac{\partial n_j}{\partial t} +\frac{\partial}{\partial x} (n_j{v}_j)=0,\label{eq2}
 \end{equation}
 where $d/dt\equiv\partial/\partial t+ {v}_j \partial/\partial x$ and we have assumed that the time scale of variation of the pressure is much smaller than that of the electron and positron density fluctuations, i.e., $(1/P_j)(dP_j/dt)\gg (1/n_j)(dn_j/dt)$. The symbols  $q_j$,  $n_j$, $v_j$,   $P_j$, and $\gamma_j=1/\sqrt{1-v_j^2/c^2}$, respectively, denote the particle's charge, the  fluid   number density in the laboratory frame such that $n_j/\gamma_j$ is the proper number density,  the fluid velocity,   the total relativistic degeneracy pressure at  finite temperature, and the Lorentz factor for the $j$-th species particles [$j=e~(p)$ for electrons (positrons)]. Also, $\phi$ is the electrostatic potential, $q_j=-e~(e)$ for electrons (positrons) with $e$ denoting the elementary charge, and     $H_j$ is the relativistic enthalpy per unit volume of each fluid species $j$, which includes the rest mass energy density, the internal energy density and the relativistic pressure.
 \par 
 The equations for the classical thermal ion fluids are
 \begin{equation}\label{eq-cont}
 \frac{\partial {n_i}}{\partial t }+ \frac{\partial \left(n_i v_i\right)}{\partial x } =0,
 \end{equation}
   \begin{equation}\label{eq-moment-i}
\frac{\partial v_i}{\partial t }+ v_i \frac{\partial v_i}{\partial x }=-\frac{e}{m_i}\frac{\partial \phi}{\partial x }-\frac{k_BT_i}{m_in_i}\frac{\partial  n_i}{\partial x },
 \end{equation}
 $m_i$, $n_i$, $v_i$, and $T_i$, respectively, denote the mass, number density, velocity, and the thermodynamic temperature of ions, and $k_B$ is the Boltzmann constant. 
 The  above set of equations \eqref{eq1}-\eqref{eq-moment-i} are closed by  the following Poisson equation.  
 \begin{equation}\label{eq-poisson}
 \frac{\partial^2\phi}{\partial x^2}=4\pi e (n_e-n_p-n_i).
 \end{equation}
 %%%%%%%%%%%%%%%%%%%%%%%%%%%%%%%%%%
\par 
It is imperative to make some assumptions on the fluid equations of electrons and positrons without loss of generality in the physics of ion-acoustic oscillations and normalize the physical quantities for brevity. 
For the excitation of ion-acoustic waves (IAWs), the inertial effects of relativistic electrons and positrons (with mass $m$), compared to those of ions, can be neglected due to $H_jm\ll m_i$ \cite{haas2016JPP}. This is valid for equilibrium number density, $n_{j0}\ll3.6\times10^{39}$ cm$^{-3}$. Also, because of their heavy inertia and slow time scale of oscillations,  compared to those of electrons and positrons, the ions are assumed to be classical, nonrelativistic and nondegenerate \cite{haas2016JPP}.  Furthermore, at the ionic time scale, the time variations of the electron and positron pressures can be assumed to be small, i.e., $(\gamma_j^2v_j/c^2)(\partial/\partial t)\ll(\partial/\partial x)$. This is justified since the phase velocity of ion-acoustic waves can be shown to be well below the speed of light $c$ in vacuum (See Sec. \ref{sec-linear} for clarification)  and the particle velocity  $v_j$ does not exceed $c$.  Thus,   Eqs. \eqref{eq1}-\eqref{eq-poisson} reduce, in dimensionless forms, to 
%%%%%%%%%%%%%%%%%%%%%%%%%%%%%%%%%%%%%%%%%%%%%%%%%%%%%%%%%%%%%%%%%%%%%%%%%%%%
 \begin{equation}\label{eq-e-moment}
 0=\frac{\partial\phi}{\partial x} -\frac{1}{n_e} \frac{\partial  {P}_e  }{\partial x },  
 \end{equation}
 \begin{equation}\label{eq-p-moment}
 0=\frac{\partial\phi}{\partial x}+\frac{\sigma_p}{ n_p} \frac{\partial  {P}_p  }{\partial x },
 \end{equation} 
 \begin{equation}\label{eq-cont-ion}
 \frac{\partial {n_i}}{\partial t }+ \frac{\partial \left(n_i v_i\right)}{\partial x } =0,
 \end{equation}
 \begin{equation}\label{eq-moment-ion}
\frac{\partial v_i}{\partial t }+ v_i \frac{\partial v_i}{\partial x }=-\frac{1}{\nu_e}\frac{\partial \phi}{\partial x }-\frac{\sigma_i}{\nu_e n_i}\frac{\partial  n_i}{\partial x },
 \end{equation}
 \begin{equation}\label{eq-poisson-dimless}
 \frac{\partial^2\phi}{\partial x^2 }=\nu_e({\alpha_e n_e}-{\alpha_p n_p}-n_i).
 \end{equation}
In Eqs. \eqref{eq-e-moment} to \eqref{eq-poisson-dimless}, different physical quantities are normalized as follows. The number density $n_j$ is normalized by its   unperturbed value $n_{j0}$ ($j=i$ for ions, $j=e$ for electrons and $j=p$ for positrons), the electrostatic potential $\phi$ is normalized by ${k_BT_e}/{e}$,  the electron (positron) relativistic pressure ${P}_e$ (${P}_p$) is normalized by $n_{e0}K_B T_e$ ($n_{p0}K_B T_p$), and the ion fluid velocity  $v_i$ is normalized by the   ion-acoustic speed $
c_s=\sqrt{({1}/{m_i})\left( {dp_e}/{dn_e}\right)_0}$. Here, the suffix $0$ denotes the value calculated at equilibrium. Furthermore, the time $(t)$ and the space $(x)$ variables are normalized by the ion plasma period $\omega_{pi}^{-1}=\left({{4\pi n_{i0}e^2}/{m_i}}\right)^{-{1/2}}$  and the effective Debye length $\lambda_{D}$ ($= {c_{s}}/{\omega_{pi}}$)  respectively. Also,  $\sigma_i=T_i/T_e$, $\sigma_p ={T_p}/{T_e}$, $\alpha_e={1}/\left({1-\delta}\right) $, $\alpha_p={\delta}/\left({1-\delta}\right)$, and $ \delta={n_{p0}}/{n_{e0}}$ such that $\alpha_e=1+\alpha_p$ (The charge neutrality condition at equilibrium).
%%%%%%%%%%%%%%%%%%%%%%%%%%%%%%%%%%%%%%%%%%%%%%%%%%%%%%%%%%%%%%%%%%%%%%%%%%%%%%%
\par 
In the interior of stellar compact objects such  as those of neutron stars and white dwarfs, electrons and positrons can have  a relativistic speed   and an arbitrary degree of degeneracy. There are different equations of state to model the degenerate matter of these compact stars. One particular, which efficiently represents the relevant physics, especially of white dwarfs, is the Chandrasekhar equation of state at finite temperature \cite{boshkayev2016equilibrium}.  
%%%%%%%%% 
\par
We consider the following expression for the electron/positron number density that follows from the Fermi-Dirac statistics \cite{boshkayev2016equilibrium,shah2010effect}.
\begin{equation}\label{eq-nj}
n_j=\int f_jdp_j=\frac{2}{(2\pi \hbar)^3}\cdot \int_{0}^{\infty} \frac{4\pi p_j^2}{\exp  \left(\frac{ E_j(p_j)- \mu_j}{k_BT_j}\right)+1} dp_j,
\end{equation}  
where $\hbar$ is the Planck's constant divided by $2\pi$ and $\mu_j$ is the  chemical potential energy for electrons and positrons without the rest mass energy.  Also, $E_j(p_j)=\sqrt{c^2p_j^2+m^2c^4}$  is the relativistic  energy and $p_j$ the relativistic momentum of $j$-th species particle. 
\par
Equation \eqref{eq-nj} can be put into the  following alternative form  \cite{boshkayev2016equilibrium,timmes1999} 
\begin{equation}\label{eq-nj1}
\begin{split} 
 n_j=\frac{8\pi\sqrt2}{(2\pi\hbar)^3}m^3c^3\beta_j^{3/2}\left[F_{1/2}(\eta_j,\beta_j)+\beta_j F_{3/2}(\eta_j,\beta_j) \right],
\end{split}
\end{equation}
where $F_k$ is the relativistic Fermi-Dirac integral of order $k$, given by,
\begin{equation}\label{eq-FD-int}
 F_{k}(\eta_j,\beta_j)=\int_{0}^{\infty}\frac{t_j^k\sqrt{1+(\beta_j/2)t_j}}{1+\exp(t_j-\eta_j)}dt_j,
 \end{equation}
in which $ \beta_j=k_BT_j/mc^2$ is the relativity parameter, $ t_j= E_j(p_j)/k_BT_j$, and  $\eta_j={\mu_j}/k_BT_j$ is the normalized chemical potential energy. 
\par 
The degeneracy pressure of the $j$-th species particle at finite temperature $(T_j\neq 0~K)$  is given by \cite{boshkayev2016equilibrium,timmes1999}
\begin{equation}\label{eq-pj0}
P_j=\frac{1}{3\pi^2 \hbar^3}\int_{0}^{\infty} \frac{p_j^3}{\exp  \left(\frac{ E_j(p_j)- \mu_j}{k_BT_j}\right)+1} dE_j,
\end{equation}
which can be expressed as
\begin{equation}\label{eq-pj1}
\begin{split}
{P}_j=\frac{2^{3/2}}{3\pi^2\hbar^3}m^4c^5\beta_j^{5/2}\left[F_{3/2}(\eta_j,\beta_j)+\frac{\beta_j}{2} F_{5/2}(\eta_j,\beta_j) \right].
\end{split}
\end{equation}
Next, using Eqs. \eqref{eq-nj1} and \eqref{eq-pj1}, it can be shown that 
\begin{equation}
 \frac{1}{n_j}\frac{\partial P_j}{\partial x}\approx \frac{\partial \mu_j}{\partial x}.
\end{equation}
 This result, when applied to the momentum balance equations for inertialess electrons and positrons, i.e.,
\begin{equation}\label{eq-moment}
 0=e\frac{\partial\phi}{\partial x} -\frac{1}{n_e} \frac{\partial  {P}_e  }{\partial x }, ~0=e\frac{\partial\phi}{\partial x}+\frac{1}{ n_p} \frac{\partial  {P}_p  }{\partial x }, 
 \end{equation}
gives  $\mu_j=-q_j\phi+\mu_{j0}$, where $\mu_{j0}$ is the value of $\mu_j$ at $\phi=0$. So, we must  replace $\mu_j$   by $-q_j\phi+\mu_{j0}$   ( or $-q_j\phi+\mu_{j}$  for brevity, keeping in mind  that $\mu_j$ is now the value at $\phi=0$)   in Eqs. \eqref{eq-nj1} and \eqref{eq-pj1} to obtain the following modified expressions for the density $n_j$ and the pressure $P_j$.  
\begin{equation}\label{eq-nj22}
\begin{split} 
 n_j=\frac{8\pi\sqrt2}{(2\pi\hbar)^3}m^3c^3\beta_j^{3/2}\left[F_{1/2}(\tilde{\eta}_j,\beta_j)+\beta_j F_{3/2}(\tilde{\eta}_j,\beta_j) \right],
\end{split}
\end{equation}
 \begin{equation}\label{eq-Pj22}
\begin{split}
{P}_j=\frac{2^{3/2}}{3\pi^2\hbar^3}m^4c^5\beta_j^{5/2}\left[F_{3/2}(\tilde{\eta}_j,\beta_j)+\frac{\beta_j}{2} F_{5/2}(\tilde{\eta}_j,\beta_j) \right],
\end{split}
\end{equation}
where $\tilde{\eta}_j=(\mu_j-q_j\phi)/k_BT_j$ is the normalized electrochemical potential energy. The energy $E_j$, appearing in Eq. \eqref{eq-nj}, is now modified to $\epsilon_j\equiv E_j+q_j\phi_j$, implying that  both the free and trapped particles are to be taken into account in the potential well. Here, particles having $\epsilon_j>0$ and $\epsilon_j<0$ are referred as the free and trapped particles  respectively, and the trapping   occurs for $\epsilon_j=0$ \cite{shah2010effect}.  
%%%%%%%%%%%%%%%%%%%%%%
%%%%%%%%%%%%%%%%%%%%%%%%%%%%%%%%
Since $\mu_j$ has the rest mass energy of electrons or positrons removed, it is the kinetic chemical potential for which the rest mass appears explicitly in the  positron chemical potential, i.e., $\mu_p=-\mu_e-2mc^2$, which gives
\begin{equation}
\xi_e=-\sigma_p\xi_p-2/\beta_e,~\rm{and}~\eta_e=-\sigma_p\eta_p-2/\beta_e,
\end{equation} 
where $\xi_j=\mu_{j0}/k_B T_j\equiv \mu_{j}/k_B T_j$ (since we have replaced $\mu_{j0}$ by $\mu_j$ for brevity) is the degeneracy parameter for electrons $(j=e)$ and positrons $(j=p)$ at equilibrium (i.e., the value of $\tilde{\eta}_j$ at $\phi=0$). Furthermore, $\xi_j$ satisfies the following harsh condition at zero relativistic and zero electrostatic potential energies $(\epsilon_j=0)$:
\begin{equation}
\sum_{j=e,p}\left[1+\exp\left(-\xi_j\right) \right]^{-1}\leq1.
\end{equation}
%%%%%%%%%%%%%%%%%%%%%%%%%%%%%%%%%%%%%%%%%%%%%%%%%%%%%%%%%%%%%%%%%%%%%%%%%%%%%  
\par 
Typically, in the non-relativistic regime of a Fermi gas at finite temperature, $\beta_j\equiv k_BT_j/mc^2\ll1$, whereas in the ultra-relativistic regime, we have   $\beta_j\gg1$. However, these limiting cases have been considered in the literature but in nonrelativistic electron-ion plasmas \cite{shah2011}. So, we are  interested in the intermediate regime  in which the particle's Fermi energy and the thermal energy do not differ significantly, i.e., $T_{Fj}>T_j$ and the particle's thermal energy is close to the rest mass energy, i.e., either $\beta_j<1$ or $\beta_j>1$. 
\par 
Thus,  evaluating the integrals $F_k(\tilde{\eta}_j,\beta_j)$ in Eq. \eqref{eq-nj22} and following   the method by Landau and Lifshitz \cite{landau2013statistical}, we obtain the following expression for the number densities of electrons and positrons (in dimensionless forms) in two different cases of $\beta_j<1$ and $\beta_j>1$.
% \begin{widetext}
\begin{equation} \label{eq-nj2}
n_j=
\begin{cases}
\begin{split}
 &A_j\left[ \left\lbrace\left(1+\phi_j/\xi_j\right)^{3/2}+\frac{\pi^2}{8}\left(1+\phi_j/\xi_j\right)^{-1/2}\xi_j^{-2} \right. +\frac{7\pi^4}{640}\left(1+\phi_j/\xi_j\right)^{-5/2}\xi_j^{-4}\right\rbrace+\frac{\xi_j \beta_j}{2}\left\lbrace\left(1+\phi_j/\xi_j\right)^{5/2}\right.& \\
   &\left. \left. +\frac{5\pi^2}{8}\left(1+\phi_j/\xi_j\right)^{1/2}\xi_j^{-2} -\frac{7\pi^4}{384}\left(1+\phi_j/\xi_j\right)^{-3/2}\xi_j^{-4}\right\rbrace \right], & \text{for $\beta_j<1$}\\ \\
 &A_j\left[\Bigl\{\left(1+\phi_j/\xi_j\right)^2+\frac{\pi^2}{3}\xi_j^{-2} \Bigl\} +\frac{\beta_j \xi_j}{2}\Bigl\{\left(1+\phi_j/\xi_j\right)^3 +\pi^2 \xi_j^{-2}\left(1+\phi_j/\xi_j\right)\Bigl\} \right],& \text{for $\beta_j>1$}.
\end{split}
\end{cases}
\end{equation}
where the coefficient $A_j$ is given by
\begin{equation} \label{expression of A}
A_j=
\begin{cases}
\begin{split}
&\left[\left(1+\frac{\pi^2}{8}\xi_j^{-2}+\frac{7\pi^4}{640}\xi_j^{-4}\right)+\frac{\xi_j \beta_j}{2}\left(1+\frac{5\pi^2}{8}\xi_j^{-2}-\frac{7\pi^4}{384}\xi_j^{-4}\right)\right]^{-1},  & \text{for $\beta_j<1$}, \\ \\
 &\left[\left(1+\frac{\pi^2}{3}\xi_j^{-2}\right) +\frac{\beta_j \xi_j}{2}\left(1+\pi^2\xi_j^{-2}\right)\right]^{-1},& \text{for $\beta_j>1$}.
\end{split}
\end{cases}
\end{equation}
%\end{widetext} 
 Also,  $\phi_e=\phi$,  $\phi_p=-\phi/\sigma_p$, and we have assumed $\eta_j>1$. The validity of this restriction of $\eta_j$ will be justified later in Sec. \ref{sec-phys-regime}.
   Thus, we have the generalized expression for the ion-acoustic speed as
    \begin{equation}
    c_s\equiv\sqrt{\frac{1}{m_i}\left(\frac{dp_e}{dn_e}\right)_0}=\sqrt{\frac{\nu_e K_BT_e}{m_i}},
    \end{equation}

where
%\begin{widetext}
\begin{equation}
\nu_e=
\begin{cases}
\begin{split}
&\frac{2\xi_e}{3 A_e} \left[ \left(1-\frac{\pi^2 }{24}\xi_e^{-2}-\frac{7\pi^4 }{384}\xi_e^{-4}\right)+\xi_e \beta_e\left(1+\frac{\pi^2 }{8}\xi_e^{-2}+\frac{7\pi^4 }{640}\xi_e^{-4}\right) \right]^{-1}, & \text{for $\beta_{e}<1$} \\ \\
&\frac{2\xi_e}{3 A_e} \left[1+\beta_e\xi_e\left(1+\frac{\pi^2}{3}\xi_e^{-2}\right)\right]^{-1}, & \text{for $\beta_{e}>1$}.
\end{split}
\end{cases}
\end{equation}
%\end{widetext}
%%%%%%%%%%%%%%%%%%%%%%%%%%%%%
\par 
It may be necessary to compare the new expression of the number density $n_j$ [Eq. \eqref{eq-nj2}]  with that in the work of Rasheed \textit{et al.} \cite{rasheed2011}.  In the latter, the authors investigated the characteristic of ion-acoustic solitons in an electron-positron-ion plasma with nonrelativistic flow of degenerate  electrons and positrons and classical cold ions. However,   they considered the ultra-relativistic degeneracy effect  at a very low temperature (in comparison with the Fermi temperature) and assumed the chemical energy to be equal to the Fermi energy. Such an assumption may be valid for a fully degenerate plasma (at zero-temperature, or $T_j\ll T_{Fj}$) but not for plasmas with finite temperature degeneracy.  We have considered this issue in the present investigation and thereby generalized and advanced the work  of  Rasheed \textit{et al.} \cite{rasheed2011}  by considering the relativistic flow of both degenerate electrons  and positrons at finite temperature and warm classical, nonrelativistic, nondegenerate ions. With  this assumption, the electron and positron distributions [Eq. \eqref{eq-nj2}] are significantly modified by the relativistic momentum and energy as well as the finite temperature effects on the chemical energy. The latter can no longer be approximated as the  Fermi energy as in Ref. \cite{rasheed2011}. Nevertheless, the expression of the number density in Ref. \cite{rasheed2011} can be recovered  in the limit of $\beta_j\gg1$ and with a replacement of the chemical energy by the Fermi energy.  
%%%%%%%%%%%%%%%%%%%%%%%%%%%%%%%%%%%%
\par
We note that the degeneracy parameters $ \xi_e $ and $ \xi_p $ are related by 
 $\xi_e=-\sigma_p \xi_p-2/\beta_e$ and the relativistic parameters $\beta_e$ and $\beta_p$ are related by $\beta_p=\beta_e \sigma_p$.  While $\beta_j$ can have any value smaller or larger than unity, the values of $\xi_j$ can be obtained by using Eq. \eqref{eq-nj1}   and the charge neutrality condition   $\alpha_e=1+\alpha_p$ at $\phi=0$. In the case of $\beta_j<1$,  an expression of $\xi_e$ can be obtained as 
 \begin{equation}
 \xi_{e}\approx \sqrt{\tau_e^2-\pi^2/8},
 \end{equation}
   where $\tau_e=T_{Fe}/T_e$. However, an explicit expression of $\xi_e$ for the case of $\beta_j>1$ can not be obtained in a straightforward way. So, we will use some approximate results for $\xi_j$ that were obtained in different contexts \cite{thomas2020}. Moreover, the normalized chemical potential $\xi_j$ can assume from large negative to large positive values. For example, in metallic plasmas, it has been shown that both the Thomas-Fermi (TF) model and the ideal free electron gas (IFEG) model predict approximately the same results for the electron chemical potential \cite{shi2014}. Given an   electron mass density $n_{e0}\sim 0.5$ gm cm$^{-3}$, as the thermal energy reduces from $5\times10^2$ ev to $0.1$ ev, the chemical potential  $\xi_e$ increases from $-5$ to more or less $20$, i.e.,   $0\lesssim\xi_e\lesssim20$ for $0.1\lesssim T~\rm(ev)\lesssim1.4$ and $-5\lesssim\xi_e\lesssim0$ for $1.4\lesssim T~\rm(ev)\lesssim10^2$. We, however, assume that at finite temperature,  the electrons and positrons   have energy states in between $k_BT_j$ and $k_BT_{Fj}$. So, negative values  of $\xi_e$ may not be admissible, otherwise one can have $T_{Fe}<T_e$.   
     In astrophysical environments, it has been found that as the particle temperature drops from $24$ keV to $12$ keV, the normalized electron chemical potential increases  from $0.01$ to $10$ and reaches a steady state value \cite{thomas2020}.  
   In particular, in the  limit of full degeneracy $(T_{Fj}\gg T_j)$, $\mu_e\approx k_B T_{Fe}=\hbar^2(3\pi^2n_{e0})^{2/3}/2m$, so that one obtains $\nu_e\approx (2/3)\tau_e$, $c_{s}\approx\sqrt{\left(2/3\right) k_BT_{Fe}/{m_i}}$, $\lambda_{D}\approx\sqrt{ \left(2/3\right)k_B T_{Fe}/{m_i \omega_{pi}^2}}$, and the Fermi pressure law, $\mathcal{P}_j=(2/5)n_{j0}E_{F_j}(n_j/n_{j0})^{5/3}$, where $E_{F_j}\equiv k_B T_{Fj}$,  i.e., the well-known  results for fully degenerate plasmas are retrieved. 
  %%%%%%%%%%%%%%%%%%%%%%%%
  \section{Physical regimes of ion-acoustic waves}\label{sec-phys-regime}
%%%%%%%%%%%%%%%%%%%%%%%%%%%%%%%%%%%%%%
In the previous section  \ref{sec-model}, we have described the basic fluid model for the excitation of ion-acoustic waves.   Nevertheless, it is pertinent to discuss the validity domains of the general theory as well as the existence domains of small amplitude ion-acoustic waves and solitons, to be investigated in Secs. \ref{sec-linear} and \ref{sec-nonlinear}. It is also highly demanding to identify precisely the key physical parameters and their regimes where the linear wave mode and the nonlinear excitation  of ion-acoustic solitons can be looked for.  Clearly, the theory is more applicable to intermediate regimes of high density and moderate temperature plasmas where 
(i) the particle's thermal energy is close to the rest mass energy, i.e.,  $\beta_j\equiv k_BT_j/mc^2\sim1$ (So, either $\beta_j<1$ or $\beta_j>1$) and (ii) the thermal and Fermi energies of electrons and positrons  do not  differ significantly, i.e.,   $T_j/T_{Fj}\lesssim1$  ($j=e,~p$ stand for electrons and positrons). The case with  $T_j/T_{Fj}>1$ is not admissible as we have assumed the electrons and positrons to have energy states in between $k_BT_j$ and $k_BT_{Fj}$ at finite temperature. Also, in this case, the electron chemical potential may be negative which may violate our assumption of $\eta_j>1$.
\par 
 In the fluid model, the electron/positron inertia has been neglected due to $H_jm\ll m_i$. This assumption is justified if the particle number density is well below   the critical density $3.6\times10^{39}~\rm{cm}^{-3}$ and the particle temperature is not significantly high, i.e.,  $T_j\lesssim10^{10}$ K.  Also, we have safely neglected the time variation of the relativistic Fermi pressure, because even in plasmas with relativistic flow of electrons and positrons, the phase velocity of ion-acoustic waves should remain well below the speed of light in vacuum $c$. This will be clarified in Sec. \ref{sec-linear}. 
%%%%%%%%%%%%%
\par 
Relativistic,  multi-component, astrophysical plasmas can occur in a wide variety of high-energy-emitting objects like white dwarfs, neutron stars, black holes, and active galactic nuclei (AGNs). In these environments,   the particle distribution  strongly depends on the various physical processes including the pair creation and annihilation. The latter can largely occur for plasmas in local thermal and chemical equilibrium. It has been shown  that such annihilation rate can be of the order of  $10^{15}~\rm{s}^{-1}$ near  $T_j=10^4~\rm{KeV}\approx1.16\times10^{11}$ K. However, at these situations it significantly drops with decreasing values of $T_j$ and freezes out at $T_j=16~\rm{KeV}\approx1.85\times10^8$ K \cite{thomas2020}. In the core of white dwarfs with central temperature $T_j\sim10^7-10^{10}~\rm{K}$   and particle number density $\sim10^{28}-10^{35}~\rm{cm}^{-3}$, the ion plasma frequency ($\sim10^{17}~\rm{s}^{-1}$ for $n_{j0}\sim10^{28}~\rm{cm}^{-3}$) can still be much higher than the annihilation rate ($\sim10^{15}~\rm{s}^{-1}$), i.e., the electron-positron annihilation time in high-density regimes can remain  longer than the ion plasma period. So, for the excitation of ion-acoustic waves in high density plasmas,   the electron-positron annihilation can be safely neglected. 
\par 
On the other hand,  in astrophysical environments,  when  the temperatures of electrons and positrons drop below
$10^9$ K but still higher than $10^7$ K, the electrons and positrons may not be in thermal and chemical equilibrium. However, since they can even strongly scatter with the plasma, their distributions can still be  the Fermi-Dirac as described in Eq. \eqref{eq-FD-int}.   Such a deviation from the chemical equilibrium implies that the electron and positron degeneracy parameters evolve separately and hence the appearance of different $\xi_e$ and $\xi_p$. It has been found that as the temperature reduces below $10^9$ K, the degeneracy parameter $\xi_j\equiv \mu_j/k_BT_j$ increases from $10^{-2}$, but reaches a steady state value $10$ at a smaller value of $T_j$. So, the values of $\xi_j>1$ are reasonably good as we have assumed $\eta_j>1$ for the expansion of the normalized densities $n_j$ [Eq. \eqref{eq-nj2}].  Also, at this nonequilibrium state, the positron to electron density ratio $\delta$ drops  below the unity \cite{thomas2020}. This is also justified from the charge neutrality condition at equilibrium $\alpha_e=1+\alpha_p$.  Furthermore, in most astrophysical plasma environments \cite{timmes1999}, the electron and positron temperatures do not differ significantly and the ion temperature is typically low compared to that of electrons or positrons, i.e.,  $\sigma_p\sim1$ and $\sigma_i<1$.
\par It is to be mentioned that although the conditions $T_j/T_{Fj}\lesssim1$ and $\beta_j\sim1$ may be fulfilled in the laser fusion experiments, e.g., at the National Ignition Facility (NIF) with number density $\sim10^{25}$  cm$^{-3}$ \cite{hurricane2014}, as well as in electrical explosion of metal wires with mass density $10^{23}$   cm$^{-3}$ \cite{shi2014}, the electron-positron annihilation rate in these environments can no longer be negligible compared to the ion plasma oscillation frequency. Thus, the present plasma model can be more relevant in the environments of white dwarfs. Before we specify the parameter regimes, the   plasma parameters responsible for the description of ion-acoustic waves can be identified as  $\beta_j$, $\xi_j$, $\sigma_i$, $\sigma_p$, and $\delta$. Since each pair of $(\beta_e,~\beta_p)$ and $(\xi_e,~\xi_p)$ are connected to each other by  a couple of relations as mentioned in Sec. \ref{sec-model}, the key parameters are $\beta_e$, $\xi_e$, $\sigma_i$, $\sigma_p$, and $\delta$. Thus,  the parameter regimes for the existence of ion-acoustic wave mode and ion-acoustic solitons may be classified in two cases as follows: 
\begin{itemize}
\item {\bf Case I, $\beta_j<1$:} The condition $\beta_j<1$ is fulfilled for   $T_j\lesssim5\times10^9$ K. Also,   $T_{Fj}>T_j$ holds for $n_{j0}\gtrsim2\times10^{30}$  cm$^{-3}$ and $T_j\lesssim5\times10^8$ K. However, since as per our assumption, the Fermi energy and the thermal energy do not differ  significantly, we consider the range for the density as  $n_{j0}\sim2\times10^{30}-10^{32}$  cm$^{-3}$ for a fixed temperature $T_j\sim5\times10^9$ K and the range for the temperature as $T_j\sim6\times10^8-5\times10^9$ K for a fixed number density $n_{j0}\sim2\times10^{30}$  cm$^{-3}$. The normalized chemical potential varies in the interval  $1<\xi_j\lesssim10$.    \\
\item {\bf Case II, $\beta_j>1$:} The condition $\beta_j>1$ is fulfilled in the temperature regime, $T_j\sim6\times\left(10^9-10^{10}\right)$ K for a fixed number density $n_{j0}\sim2\times10^{30}$ and in the number density regime, $n_{j0}\sim2\times\left(10^{30}-10^{32}\right)$  cm$^{-3}$ for a fixed temperature $T_j\sim6\times10^9$ K. The  chemical potential varies as in  Case I. 
\end{itemize} 
  %%%%%%%%%%%%%%%%%%%%%%%
  \par 
  In the following Secs. \ref{sec-linear} and \ref{sec-nonlinear}, we will study the linear and nonlinear theory of ion-acoustic waves in relativistic degenerate e-p-i plasmas. Specifically, we will focus on the two parameter regimes as defined before to establish the existence   of  ion-acoustic wave modes and ion-acoustic solitons. The properties of these solitons  will also be studied with the variation of parameters.  
\section{Linear Analysis} \label{sec-linear}
We  consider the propagation of electrostatic waves in relativistic degenerate e-p-i plasmas in the limit of small amplitude perturbations  for which  any nonlinear effects can be neglected  and look for the existence and the characteristics of the ion-acoustic mode through a linear dispersion relation.  In order to obtain this dispersion relation for IAWs, we linearize   Eqs.  \eqref{eq-e-moment}-\eqref{eq-poisson-dimless} by considering the dependent variables as a sum of their equilibrium and perturbation parts, i.e.,
$n_j=1+n_{j1}$,   $ v_i= 0+v_{i1}$, $\phi=0+\phi_1$ etc. Next, we assume the perturbed (with suffix $1$) quantities to vary as plane waves with the   wave number $k$ (normalized by $\lambda_{D}^{-1}$) and the wave frequency $\omega$ (normalized by $\omega_{pi}$) of the form  $\sim\exp (ik x -i\omega t)$. Thus, we obtain the following  dispersion relation for IAWs.
\begin{equation}\label{eq-disp-rel}
\frac{\omega^2}{k^2}= \frac{1}{\Lambda+k^2}+ \frac{\sigma_i}{\nu_e},
\end{equation}
where $\Lambda=\nu_e\left(\alpha_e a_{0e}+\alpha_p a_{0p}/\sigma_p\right)$ and $a_{0j}$ (for $j=e,~p$) is given by
%\begin{widetext}
\begin{equation} \label{expression of b0j}
a_{0j}=
\begin{cases}
\begin{split}
 &A_j \left[\left(\frac{3}{2}\xi_j^{-1}-\frac{\pi^2}{16}\xi_j^{-3}-\frac{7\pi^4}{256}\xi_j^{-5}\right) +\frac{\xi_j \beta_j}{2}\left(\frac{5}{2}\xi_j^{-1}+\frac{5\pi^2}{16}\xi_j^{-3}+\frac{7\pi^4}{256}\xi_j^{-5}\right)\right],  & \text{for $\beta_j<1$} \\ \\
 &A_j\left[2\xi_j^{-1}+\frac{\beta_j}{2}\left(3+\pi^2\xi_j^{-2}\right)\right].& \text{for $\beta_j>1$}.
\end{split}
\end{cases}
\end{equation}
%\end{widetext} 
\par 
From the expression of $\omega^2/k^2$ [Eq. \eqref{eq-disp-rel}] some important consequences are to be  noted as follows.  
\begin{itemize}
\item[•] The dispersion relation is a generalization of that obtained in Ref. \cite{rasheed2011} with the effects of the relativity parameters $\beta_j$ and $\xi_j$ and the thermal effects of ions ($\sigma_i$). In contrast to the work \cite{rasheed2011}, the chemical energy $\mu_j$ is no longer the Fermi energy but it may vary with the temperature $T_j$  \cite{thomas2020,shi2014}.  One can recover the dispersion equation in Ref. \cite{rasheed2011} by setting $\beta_j\gg1$, $\mu_j=k_BT_{Fj}$ and $\sigma_i=0$.    
\item[•] It is evident from the first term that the effective charge screening length is given by $\lambda_D^\text{eff}=\Lambda^{-1/2}$, which is clearly reduced due to the   positron concentration in plasmas and   decreasing values of $\beta_e~(<1)$. Physically,  as the positron is introduced into the cloud of electrons and ions, more electrons will get attracted into the Debye sphere to neutralize the charge. As a result, the plasma will be more dense and the Debye sphere will be more compact. So,  the characteristic distance over which electrons and ions were initially (before positrons are introduced) separated will be reduced. On the other hand, when electrons have low  thermal energy or $\beta_e<1$, their  possibility  to escape from the plasma cloud also reduces and as a result, the characteristic distance between the plasma particles reduces.   However, an opposite behavior occurs for $\beta_e>1$, i.e., when the particle thermal energy exceeds the rest mass energy, the screening length  gets enhanced.  
\item[•] Typically, for some   parameter values   satisfying $0<\delta<1$, $\sigma_i<1$, $\sigma_p\sim1$, $\xi_e>1$, and $\beta_e<1$ or $\beta_e>1$ as stated in Case I and Case II, we have $\Lambda\lesssim1$. So, in the long-wavelength limit, i.e., $k^2\ll\Lambda\lesssim1$, the phase velocity of IAWs approaches a constant value, i.e., 
\begin{equation}\label{eq-phase1}
\frac{\omega}{k}\approx \sqrt{ \frac{1}{\Lambda}+ \frac{\sigma_i}{\nu_e}}.
\end{equation} 
This is a prerequisite for the existence of KdV solitons.
%%%%%%%%%%%%%%%%%%%%%%%%%%%%%%
\begin{figure}
\centering
\includegraphics[width=4.5in,height=2.0in]{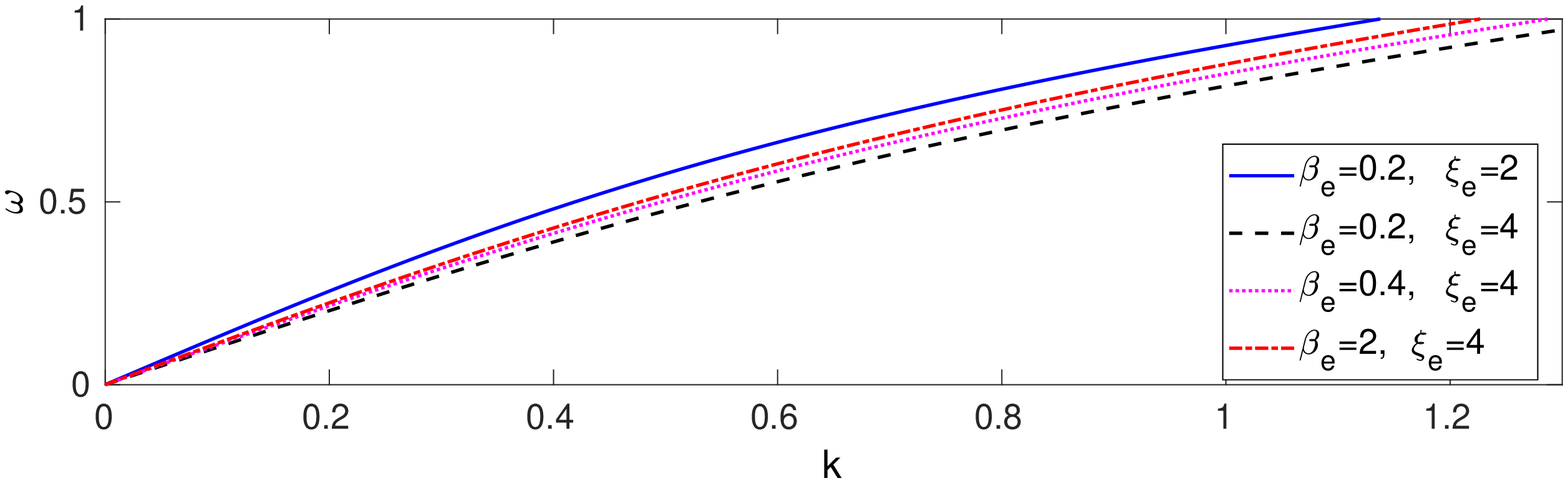}
\caption{The dispersion relation [Eq. \eqref{eq-disp-rel}] is plotted to show the variation of the wave frequency ($\omega$) against the wave number ($k$)  for different values of $\beta_e$ and $\xi_e$ as in the legend. The fixed parameter values are $\delta=0.7$, $\sigma_p=0.8$, and $\sigma_i=0.5$.}
\label{fig:disp1}
\end{figure}
%%%%%%%%%%%%%%%%%%%%%%%%%%%
Furthermore, the two terms proportional to   $a_{0e}$ and $a_{0p}$ in the expression of $\Lambda$ appear  due to the   effects of the relativistic finite temperature degeneracy of electrons and positrons. So, it follows that the phase velocity is  strongly influenced by the positron concentration, the ion  temperature, and the relativistic degeneracy of electrons and positrons. It can be shown that the phase velocity lies in the interval (in terms of its original dimension) $v_{ti}<\omega/k<\sqrt{c_s^2+v_{ti}^2}$, where $v_{ti}=\sqrt{k_BT_i/m_i}$ is the ion thermal speed. Typically, $c_s$ is larger than $v_{ti}$ and it scales as $c_s\sim\sqrt{k_BT_e/m_ec^2}\sqrt{m_e/m_i}$. If $k_BT_e\sim m_ec^2$ and $m_e\ll m_i$, we have $\omega/k<c_s\ll c$. This is the condition which we have presumed in Sec. \ref{sec-model} for the time variation of the relativistic pressure to be negligible. \\
\item[•]
In particular, when the positron contribution is dropped and ions are assumed to be cold, one can obtain the following  dispersion relation for ion-acoustic waves in relativistic degenerate electron-ion plasmas.  
\begin{equation}\label{eq-disp-reduced}
\frac{\omega^2}{k^2}= \frac{1}{a_{0e} \nu_e+k^2}.
\end{equation}
From Eq. \eqref{eq-disp-reduced}, it may be noted that  in contrast to the classical IAWs \cite{chen2012introduction}, the typical Debye screening length in relativistic degenerate electron-ion plasmas at finite temperature is also modified to $(a_{0e}\nu_e)^{-1/2}$. The latter  can also be shown to be less than or of the order of unity for typical plasma parameters as stated before.
\end{itemize}
\par    
In what follows, we  study the dispersion characteristics of IAWs by numerically solving Eq. \eqref{eq-disp-rel} for $\omega$    for different values of  $\beta_e$ and $\xi_e$ that fall within the parameter regimes defined in Case I and Case II. The dispersion curves for the IAW mode are shown in Fig. \ref{fig:disp1}. While the parameter $\beta_e$ characterizes the measure of the  thermal energy relative to the rest mass energy,   $\xi_e$ measures the degree of degeneracy of electrons and hence that of positrons. It is found that with a small increase of the value of $\beta_e$,  the  IAW frequency also increases (See the dotted and dash-dotted lines). However,  the  frequency gets significantly reduced with an increasing value of $\xi_e$ (See the solid and dashed lines). Physically, at higher thermal energies of electrons and positrons  beyond the rest mass energy, more number of wave crests  may pass a particular point (due to frequency increase)  in a given interval of time and so the ion-acoustic wave of constant amplitude may  transmit  more energy per unit time.    However, the transmission of the wave energy may be reduced when electrons and positrons approach the Fermi level with an increasing value of $\xi_e$ (See the solid and dashed lines). From Fig. \ref{fig:disp1} it is also noted that as the values of $\beta_e$ are increased or those of $\xi_e$ are decreased, the domain of  $\omega$ in terms of $k~(\lesssim1)$ for the existence of   IAW mode  reduces. In these situations, the IAWs can propagate with longer wavelengths and hence with higher energies.  On the other hand,  the effects of increasing values of $\sigma_p$ and $\sigma_i$ are to enhance a bit the wave frequency. However, such effect becomes significant in the regime of higher values of $k>1$.  The latter may   not be relevant to the study of low-frequency IAWs with longer wavelengths, because otherwise,  the ion-acoustic wavelength may become   smaller than the Debye screening length for which the plasma collective  behaviors may disappear.  
%%%%%%%%%%%%%%%%%%%%%

\section{Nonlinear evolution of ion-acoustic solitons} \label{sec-nonlinear}
%\subsection{KdV equation and its solution} 
%\label{Derivation of the KdV equation}
In this section, we will relax the extreme condition for small amplitude perturbations for which the linear theory is no longer valid and look for how  the perturbations develop into the  excitation of ion-acoustic solitary waves as the nonlinear effects intervene  the dynamics of relativistic degenerate e-p-i plasmas. Specifically, we will derive evolution equations for ion-acoustic solitons and study their properties in different parameter regimes that are defined in Case I and Case II.  
In Sec. \ref{sec-linear}, we have  seen that the dispersion properties of IAWs are distinct in these two cases. Also, in Sec. \ref{sec-model}, we have noted that the nonlinear contributions in  the electron and positron number densities [Eq. \eqref{eq-nj1}]  are significantly different  for $\beta_j<1$ and $\beta_j>1$, $j=e,~p$.  So, we will consider these two cases separately in Secs. \ref{sec-case1} and \ref{sec-case2}.  We will employ the standard reductive perturbation technique to derive the evolution equation for small amplitude ion-acoustic solitons, namely the KdV equation, and consider the critical parameter regimes where the KdV equation fails, but some other nonlinear equations like mKdV and Gardner equations describe the evolution of ion-acoustic solitons.  
\subsection{Case I, $\beta_e<1$} \label{sec-case1}
\subsubsection{KdV solitons} \label{sec-kdv}
We consider the nonlinear propagation of small-amplitude ion-acoustic perturbations and look for the evolution equation of  small-amplitude ion-acoustic solitons in relativistic degenerate plasmas at finite temperature with $\beta_j<1$, i.e., when the electron/positron thermal energy is slightly below their rest mass energy. In the weekly nonlinear theory, such an evolution equation of the KdV type can readily be obtained by using the standard reductive perturbation technique. 
   To this end, we define the   stretched coordinates using the Galilean transformation as
 \begin{equation}\label{eq-stretch}
  \xi=\epsilon^{1/2}(x-\lambda t),~ \tau=\epsilon^{3/2}t,
\end{equation} 
where $\epsilon$ ($0<\epsilon<1$) is a small expansion parameter measuring the weakness of the wave amplitudes and  $\lambda$ is the phase velocity of the IAW normalized by $c_s$. The new coordinates $\xi$ and $\tau$ are, respectively,  normalized by $\lambda_D$ and  $\omega_{pi}^{-1}$. In a general manner, one can define the stretched coordinates using the Lorentz transformation (instead of the Galilean transformation) for the relativistic fluid model as 
\begin{equation}
\xi=\epsilon^{1/2}\gamma_L\left(x-\lambda t\right),~ \tau=\epsilon^{3/2}\gamma_L\left(t-\lambda z\right),
\end{equation}
 where $\gamma_L=1/\sqrt{1-\lambda^2}$ stands for another Lorentz factor.  However, this is not necessary as the basic equations are  reduced into the forms which do not involve any relativistic Lorentz factor. Also,  the Lorentz transformation defined above would not change any qualitative features of the ion-acoustic wave dynamics. In fact, the factor $\gamma_L$  may  contribute to the  dispersion and nonlinear coefficients of the evolution equation explicitly with its different powers, which only  change their  magnitudes a bit when the ion-acoustic phase velocity $(\lambda\equiv\omega/k)$ is well below the acoustic speed $c_s$.  Furthermore,    defining the multiple scales [like Eq. \eqref{eq-stretch}] is justified, since for a small wave number (or long wavelengths) $k\sim{\cal O}(\epsilon^{1/2})$, the phase factor of a plane wave $kx-\omega t$ can be expressed by using the cold ($\sigma_i=0$, for simplicity) plasma dispersion relation $\omega=k/\sqrt{\Lambda+k^2}$ [Eq. \eqref{eq-disp-rel}] as 
\begin{equation}
kx-\omega t=\epsilon^{1/2}\left(x-\frac{1}{\sqrt{\Lambda}}t\right)+\frac{\epsilon^{3/2}}{2\Lambda^{3/2}}t 
\end{equation}
and  the phase velocity as $\lambda=1/\sqrt{\Lambda}$. The latter will be verified later.
\par    
In what follows, the dependent variables are expanded in powers of $\epsilon$ as 
 \begin{eqnarray} \label{eq-expan}
 n_j=&1+\epsilon n_{j1}+\epsilon^2
n_{j2}+\cdots,  \nonumber \\
v_i=&\epsilon v_{i1}+\epsilon^2 v_{i2}+\cdots, \\
\phi=&\epsilon \phi_1+\epsilon^2
\phi_2+\cdots. \nonumber 
 \end{eqnarray}
Next, we apply  the new coordinate transformations [Eq. \eqref{eq-stretch}] and substitute   the expansions from Eq.  \eqref{eq-expan} into Eqs. \eqref{eq-e-moment}-\eqref{eq-poisson-dimless}  and Eq. \eqref{eq-nj2}, and then equate the coefficients of different powers of $\epsilon$ from the resulting equations.
The lowest order of $\epsilon$ yields  the following expressions for the first order perturbations. 
\begin{eqnarray} \label{eq-1st-order} 
n_{i1}=\frac{1}{\lambda^2\nu_e-\sigma_i}\phi_1, \nonumber \\
n_{e1}=a_{0e}\phi_1,~ n_{p1}=-\frac{a_{0p}}{\sigma_p}\phi_1,  \\
v_{i1}=\frac{\lambda}{\lambda^2\nu_e -\sigma_i }\phi_1.  \nonumber
\end{eqnarray}
Eliminating the first-order perturbations successively and looking for their nonzero solutions, from Eq. \eqref{eq-1st-order} we obtain the following relation for the phase velocity $\lambda$.
\begin{equation}
\lambda=\left[\frac{1}{\nu_e}\left(\sigma_i+\frac{\sigma_p}{\alpha_ea_{0e}\sigma_p+\alpha_p a_{0p}}\right).\right]^{1/2}. \label{eq-lambda} 
\end{equation}
As expected, this expression of $\lambda$ exactly agrees with that of $\omega/k$, to be obtained from the dispersion equation \eqref{eq-disp-rel} in the limit of long wavelength perturbations, i.e., $k\rightarrow0$. It also justifies the consideration of $\lambda=1/\sqrt{\Lambda}$ in the coordinate transformation \eqref{eq-stretch}.
 \par 
  To the next higher order of $\epsilon$, we have 
\begin{eqnarray}
-\lambda\frac{\partial n_{i2}}{\partial \xi }+ \frac{\partial n_{i1}}{\partial \tau }+\frac{\partial}{\partial \xi}(v_{i2}+n_{i1}v_{i1})=0, \label{2nd_order_1} \\
\begin{split}
-\lambda\frac{\partial v_{i2}}{\partial \xi }+\frac{\partial v_{i1}}{\partial \tau }+ v_{i1}\frac{\partial v_{i1}}{\partial \xi }+\frac{1}{\nu_e}\frac{\partial \phi_2}{\partial \xi}+\frac{\sigma_i}{\nu_e}\left(\frac{\partial n_{i2}}{\partial \xi }-n_{i1}\frac{\partial n_{i1}}{\partial \xi }\right)=0,
\end{split} \label{2nd_order_2}\\
n_{e2}=a_{0e}\phi_2+a_{1e}\phi_1^2, ~ n_{p2}=-\frac{a_{0p}}{\sigma_p}\phi_2+\frac{a_{1p}}{\sigma_p^2}\phi_1^2, \label{2nd_order_3} \\
\frac{\partial^2\phi_1}{\partial \xi^2}=\nu_e \left(\alpha_en_{e2}-\alpha_pn_{p2}-n_{i2}\right), \label{2nd_order_4}
\end{eqnarray}
where $a_{1j}$ is given by
%\begin{widetext}
\begin{equation}\label{expression of b1j}
a_{1j}=
\begin{cases}
\begin{split}
&A_j \left[\left(\frac{3}{8}\xi_j^{-2}+\frac{3\pi^2}{64}\xi_j^{-4}+\frac{49\pi^4}{1024}\xi_j^{-6}\right) +\frac{\xi_j \beta_j}{2}\left(\frac{15}{8}\xi_j^{-2}-\frac{5\pi^2}{64}\xi_j^{-4}-\frac{35\pi^4}{1024}\xi_j^{-6}\right)\right],  & \text{for $\beta_j<1$}\\ \\
&A_j\left[\xi_j^{-2}+\frac{3\beta_j}{2}\xi_j^{-1}\right], & \text{for $\beta_j>1$}.
\end{split}
\end{cases}
\end{equation}
%\end{widetext}
Finally, eliminating all the second order quantities from Eqs. (\ref{2nd_order_1})-(\ref{2nd_order_4}) and using the results of lowest order of $\epsilon$, we obtain the following  KdV equation for the first-order electrostatic potential $\psi\equiv\phi_1$.
\begin{equation}\label{eq-kdv}
\frac{\partial \psi}{\partial \tau }+A_1B\psi\frac{\partial \psi}{\partial \xi }+B\frac{\partial^3 \psi}{\partial \xi^3 }=0,
\end{equation}
where the dispersion $(A_1B)$ and the nonlinear $(B)$  coefficients are given by
\begin{equation}
\begin{aligned} \label{eq-coeff-kdv}
A_1 =  \nu_e\left[\frac{3\lambda^2\nu_e-\sigma_i}{(\lambda^2\nu_e-\sigma_i)^3}+\frac{2\alpha_p a_{1p}}{\sigma_p^2}-2\alpha_e a_{1e}\right],~~
B =  \frac{(\lambda^2\nu_e-\sigma_i)^2}{2\lambda \nu_e^2}.
\end{aligned}
\end{equation}
Evidently both the dispersion (which causes wave broadening) and nonlinear (responsible for wave steepening) coefficients of the KdV equation \eqref{eq-kdv} are significantly modified by the contributions of the positron species, the thermal ions, as well as the relativistic degeneracy of both electrons and positrons.
\par 
 Before we analyze the characteristics of $A$ and $B$, let us first obtain a traveling wave solution of Eq. \eqref{eq-kdv}. To this end, we apply the transformation $\zeta=\xi-U \tau \equiv \epsilon^{1/2}\left[x-\left(\lambda+\epsilon U\right)t \right]$, so that $U$ is the constant speed of ion-acoustic solitons that represents a small increment of the linear phase speed of IAWs $\lambda$, and the boundary conditions, namely $\psi$, $d\psi/d\xi$, and $d^2\psi/d\xi^2\rightarrow0$ as $\xi\rightarrow\pm\infty$. Thus, we obtain  
\begin{equation} \label{eq-sol-kdv}
\psi=\psi_m \rm{sech}^2\left(\zeta/w\right),
\end{equation}
where  $\psi_m$ and $w$, respectively, denote the maximum amplitude and the width of ion-acoustic solitons, given by, 
\begin{equation} \label{eq-amp-width-kdv}
\psi_m=\frac{3U}{A_1B},~ w=\sqrt{\frac{4B}{U}},
\end{equation} 
and they are such that the relation $\psi_m w^2=12/A_1$ holds. It is to be noted that a small amplitude soliton solution, similar to that obtained in Ref. \cite{rasheed2011} by the Sagdeev pseudopotential approach,  can be recovered in the limiting case of $\beta_j\gg1$, and with the substitutions $\mu_j=k_BT_{Fj}$ and $\sigma_i=0$.   
\par
 Furthermore, the soliton energy (or soliton photon number) is given by
\begin{equation}\label{eq-sol-energy}
{\cal E}=\int_{-\infty}^{\infty} |\psi|^2d\zeta=\frac{4}{3}\psi_m^2w=\frac{24}{A_1^2}\left(\frac{U}{B}\right)^{3/2}.
\end{equation}
Since the soliton speed $U$ is directly proportional to the amplitude $\psi_m$, but inversely to the width $w$, faster
(slower) solitons may be taller (shorter) and narrower (wider). Furthermore, since the soliton energy ${\cal E}$ is directly proportional to the amplitude and width, ion-acoustic solitons with higher amplitudes (and/or widths) would evolve with higher energies in relativistic plasmas.   From the expression of $B$ in Eq. \eqref{eq-coeff-kdv}, it is evident that $B$ is always positive and   also non-zero by means of Eq. \eqref{eq-disp-rel}.  So, the KdV equation  \eqref{eq-kdv} admits  compressive or rarefactive soliton solutions according to when $A_1>0$  or $A_1<0$. However, when $A_1=0$,   the KdV equation fails to describe  the nonlinear evolution of ion-acoustic solitons. In that case, one has to look for some higher order correction terms in the  perturbation expansions, to be investigated later.
\par
\begin{figure} 
\centering
\includegraphics[width=4.5in,height=2.5in]{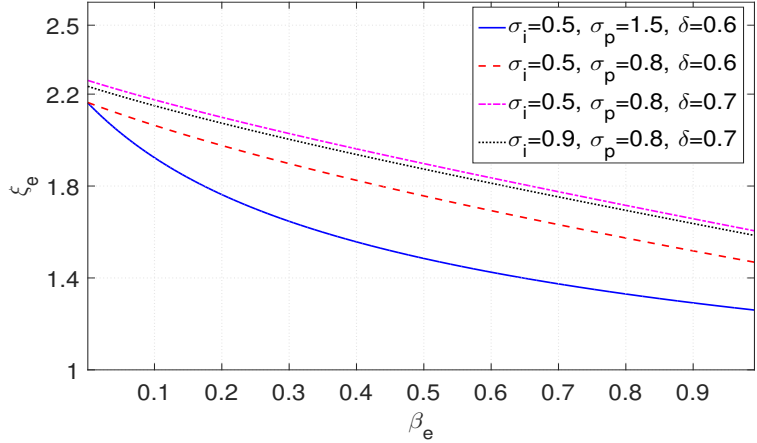}
\caption{Contour plot of $A_1=0$ in the $\beta_e-\xi_e$ plane for different values of $\sigma_i$,   $\sigma_p$, and $\delta$ as in the legend. The region above (below) the line $A_1=0$ corresponds to  the existence regime of compressive   (rarefactive) KdV solitons. For the parameter values lying on the curves of $A_1=0$ and/or  close to the curves where $A_1 \sim {\cal O}(\epsilon)$, the KdV  equation may not be valid for the evolution of small amplitude ion-acoustic solitons. }
\label{fig:A1_zero}
\end{figure}

 
We consider the parameter regimes as in Case I which involves $\beta_e<1$ and numerically investigate the properties of $A_1$ to identify  different parameter domains for which the conditions $A_1>0$, $A_1<0$, and    $A_1=0$  may be fulfilled. Figure \ref{fig:A1_zero} displays the contour plot of $A_1=0$ in the $\beta_e-\xi_e$ plane for different values of $\sigma_i$,   $\sigma_p$, and $\delta$.
%%%%%%%%%%%%%%%%%%%%%%%%%%%%%%%%%%%%%%%%%%%%%%%%%%%%%%%%%%%%%%%%%%%%%%%%%%%%%%%%%%%%% 
%%%%%%%%%%%%%%%%%%%%%%%%%%%%%%%%%%%%%%%%%%%%%%%%%%%%%%%%%%%%%%%%%
While   different points on the curves correspond to  different parameter values at which $A_1=0$, the regions above and below the curves, respectively, represent the parameter regimes where $A_1>0$ and $A_1<0$.  In the  former, the hump shaped (compressive with positive potential) solitons  may exist, whereas  in the latter, one can find dip shaped (rarefactive with negative potential) ion-acoustic solitons. Here, we call a line of $A_1=0$ as the ``critical line",  a  point $P_c\equiv (\beta_{\rm{ec}}, \xi_{\rm{ec}})$  on the critical line   as the ``critical point",   and any point $P\equiv (\beta_e, \xi_e)$ lying in the   $\beta_e-\xi_e$ plane but close  to the critical line  [i.e., in the region where $A_1 \rightarrow 0$ or $A_1 \sim {\cal O}(\epsilon)$]   as the ``close to the critical point  $P_c$".  We note that the KdV theory may not be valid for the parameter regimes at the critical points or close to the critical points. We will treat these particular cases   in Subsecs. \ref{sec-mKdV} and \ref{sec-gardner} separately.    From Fig. \ref{fig:A1_zero} it is found that as the value of the positron to electron temperature ratio $(\sigma_p)$ is reduced (See the solid and dashed lines) or that of the positron to electron density ratio $(\delta)$ is enhanced (See the dashed and dash-dotted lines),    the parameter region of rarefctive solitons corresponding to $A_1<0$ expands, while that of compressive solitons, i.e., $A_1>0$ shrinks. The influence of the ion temperature $(\sigma_i)$ on the existence regions of ion-acoustic solitons is not markedly pronounced. However, an enhancement of $\sigma_i$ expands a bit the parameter region for the compressive solitons, but reduces that of the rarefactive one (See the dotted and dash-dotted lines).  Thus, from Fig. \ref{fig:A1_zero} it may be concluded that in contrast to typical electron-ion plasmas,  within the specific domains of values of $\xi_e$ and $\beta_e$ (and so of $\xi_p$ and $\beta_p$) and for a fixed positron to electron temperature ratio,  higher the concentration of the positron species $(\delta)$ or lower the ion to electron temperature ratio $(\sigma_i)$ in relativistic degenerate e-p-i plasmas,   the more likely is the existence of rarefactive ion-acoustic solitons than the compressive solitons.   
\par
\begin{figure} 
\centering
\includegraphics[width=4.5in,height=2.0in]{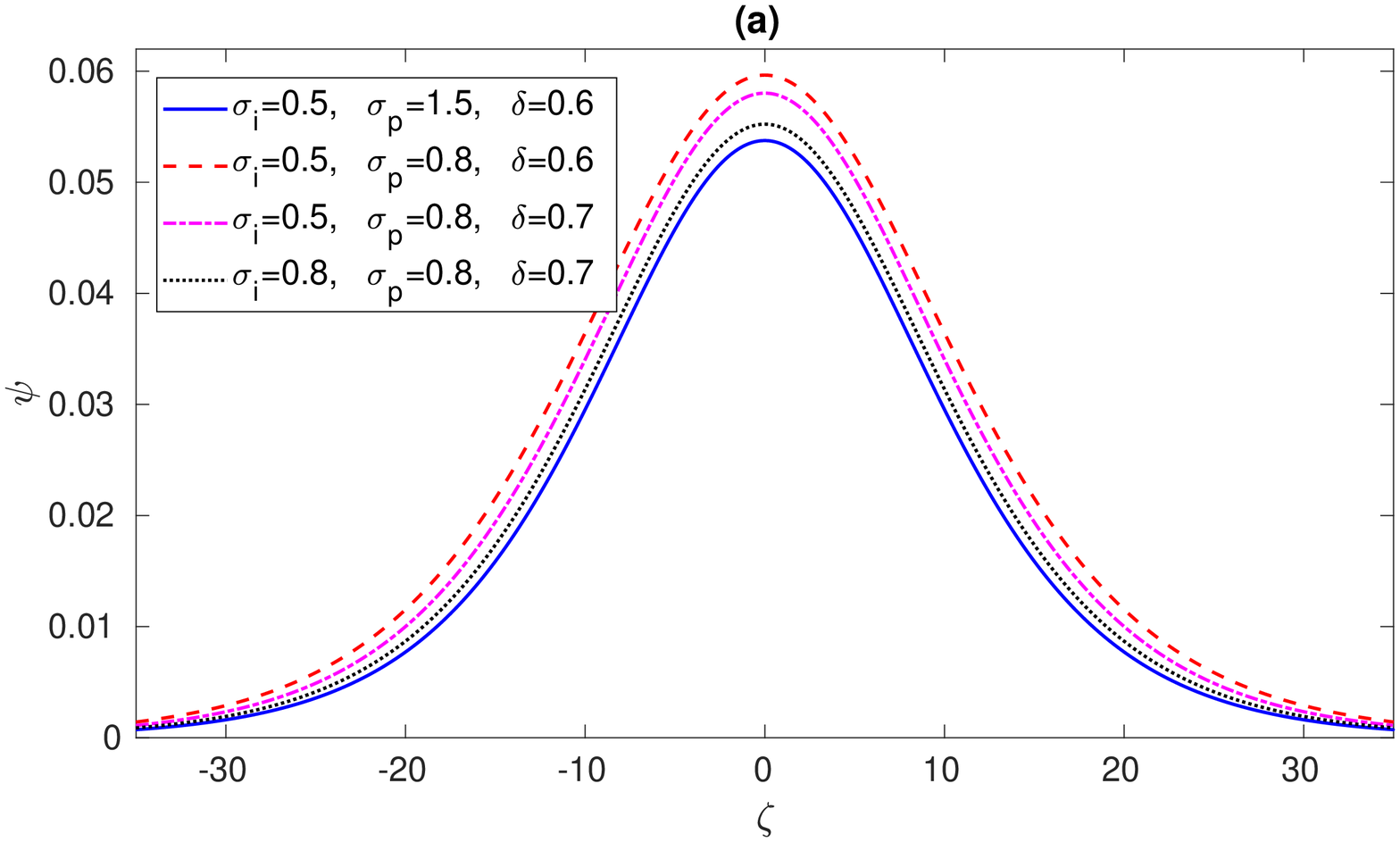}
\includegraphics[width=4.5in,height=2.0in]{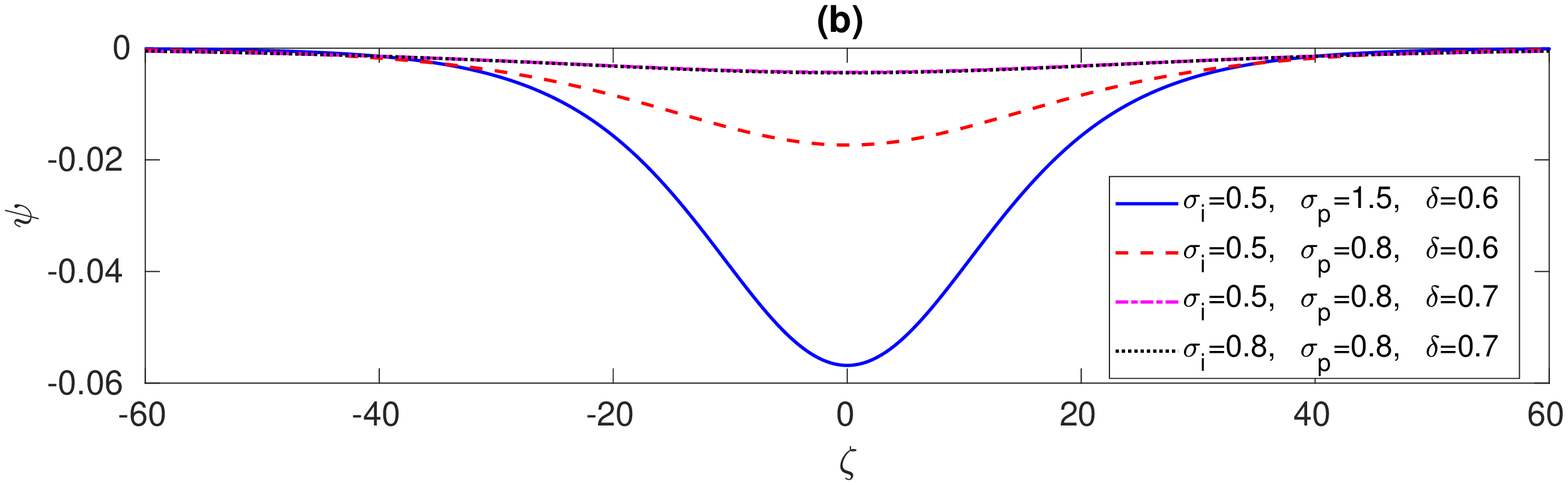}
\caption{The profiles of the compressive [subplot (a)] and rarefactive [subplot (b)] ion-acoustic solitons are shown  for different values of  $\sigma_i$,   $\sigma_p$, and $\delta$ as in the legends.  The   parameter values for the subplots (a) and (b), respectively, are $(\beta_e,\xi_e)=(0.4,5)$ and  $(\beta_e,\xi_e)=(0.1,1.8)$. Here, $U=0.01$ and the values of $\beta_e$ and $\xi_e$   are taken from the existence region (Fig. \ref{fig:A1_zero}) in  such a way that $A_1 \neq 0$ and $A_1 \nsim o(\epsilon)$.}
\label{fig:KdV_SW_sigma_delta}
\end{figure}

 
Having obtained the parameter regimes for the existence of ion-acoustic solitons away from the critical points, we plot the profiles of both the compressive and rarefactive solitons for different values of  $\sigma_i$,  $\sigma_p$, and $\delta$  as shown in Fig. \ref{fig:KdV_SW_sigma_delta}. As an illustration, for the compressive and rarefactive solitons, we consider, respectively,  the points $P \equiv (0.4,5)$ and $(0.1,1.8)$  in the $\beta_e\xi_e$ parameter space [Fig. \ref{fig:A1_zero}], which neither lie on the critical lines nor close to them, since at these points the KdV theory may not be valid.   For example, at   a critical point $P_c \simeq (0.4,1.9617)$ (or close to it) of   the dash-dotted  line of Fig. \ref{fig:A1_zero},   the amplitude of the soliton [Eq. \eqref{eq-amp-width-kdv}]  becomes extensively larger than unity, leading to the failure of the weekly nonlinear theory of small-amplitude perturbations.   In such situations, the mKdV and Gardner equations may precisely describe the evolution of ion-acoustic solitons which will be studied shortly in Secs. \ref{sec-mKdV} and \ref{sec-gardner}.    
%%%%%%%%%%%%%%%%%%%%%%%%%%%%%%%%%%%%%%%%%%%%
%%%%%%%%%%%%%%%%%%%%%%%%%%%%%%%%%%%%%%%%%%%%%%%%%%%%%%%
From Fig. \ref{fig:KdV_SW_sigma_delta}, it is found that both the widths and amplitudes of  the compressive [Subplot (a)] and rarefactive [Subplot (b)] solitons decrease with  increasing values of $\delta$  and   the reduction is significant for rarefactive solitons (See the dashed and dash-dotted lines).  Such a reduction of the amplitude and width can eventually lead to a significant decay of the soliton energy [Eq. \eqref{eq-sol-energy}].  Although the influence of ion to electron temperature ratio $\sigma_i$  on the profiles of rarefactive solitons is relatively small,  both the  amplitude and width of the compressive solitons decrease with increasing values of  $\sigma_i$ (See the dotted and dash-dotted lines). On the other hand, in contrast to the compressive solitons in which both the amplitude and width decrease, the influence of increasing  the positron to electron temperature ratio $\sigma_p$  is to increase both the   amplitude and width of the rarefactive solitons.   Thus,   it may be concluded that  the positron species (which favors the existence of rarefactive solitons, \textit{cf}. Fig. \ref{fig:A1_zero}) with thermal energies close to the electrons and with higher concentration   in plasmas can reduce the soliton energy significantly. Also, the relative influence of the plasma parameters $\sigma_p$ and $\delta$ on the profiles of the compressive and rarefactive solitons are not only different but their qualitative features also differ significantly.   
\par
\begin{figure} 
\centering
\includegraphics[width=3.2in,height=2.0in]{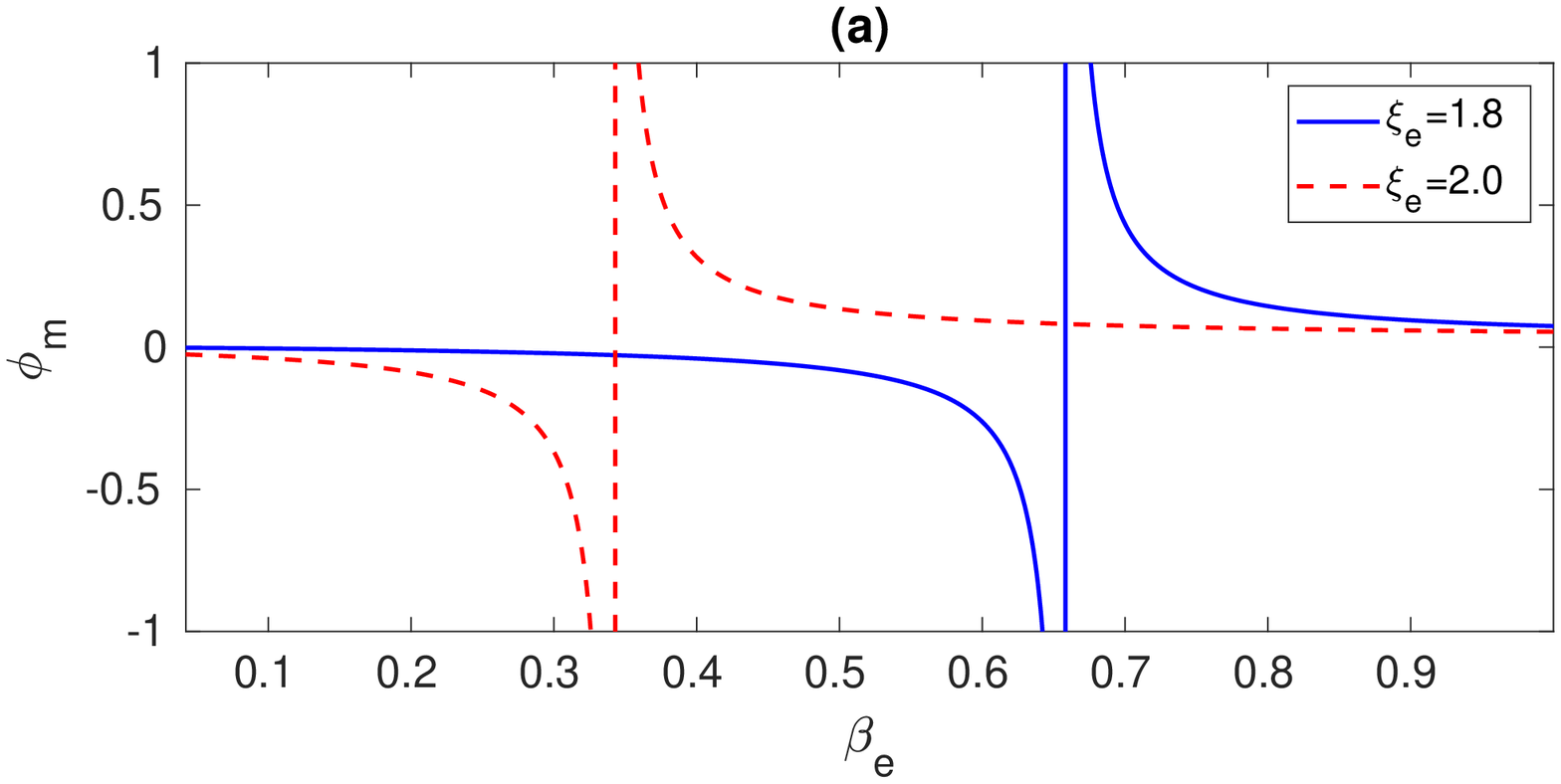}
\includegraphics[width=3.2in,height=2.0in]{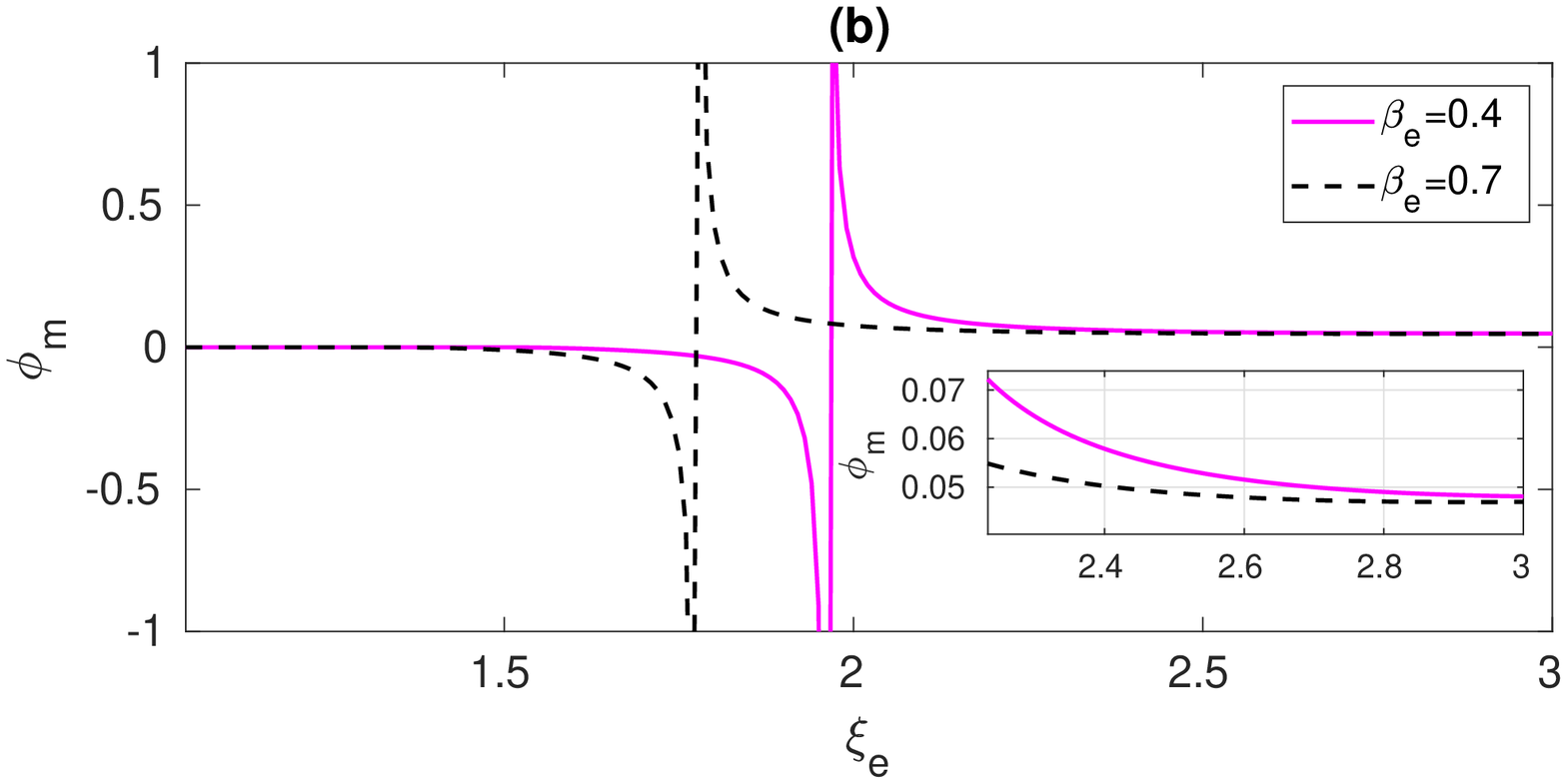}
\includegraphics[width=3.2in,height=2.0in]{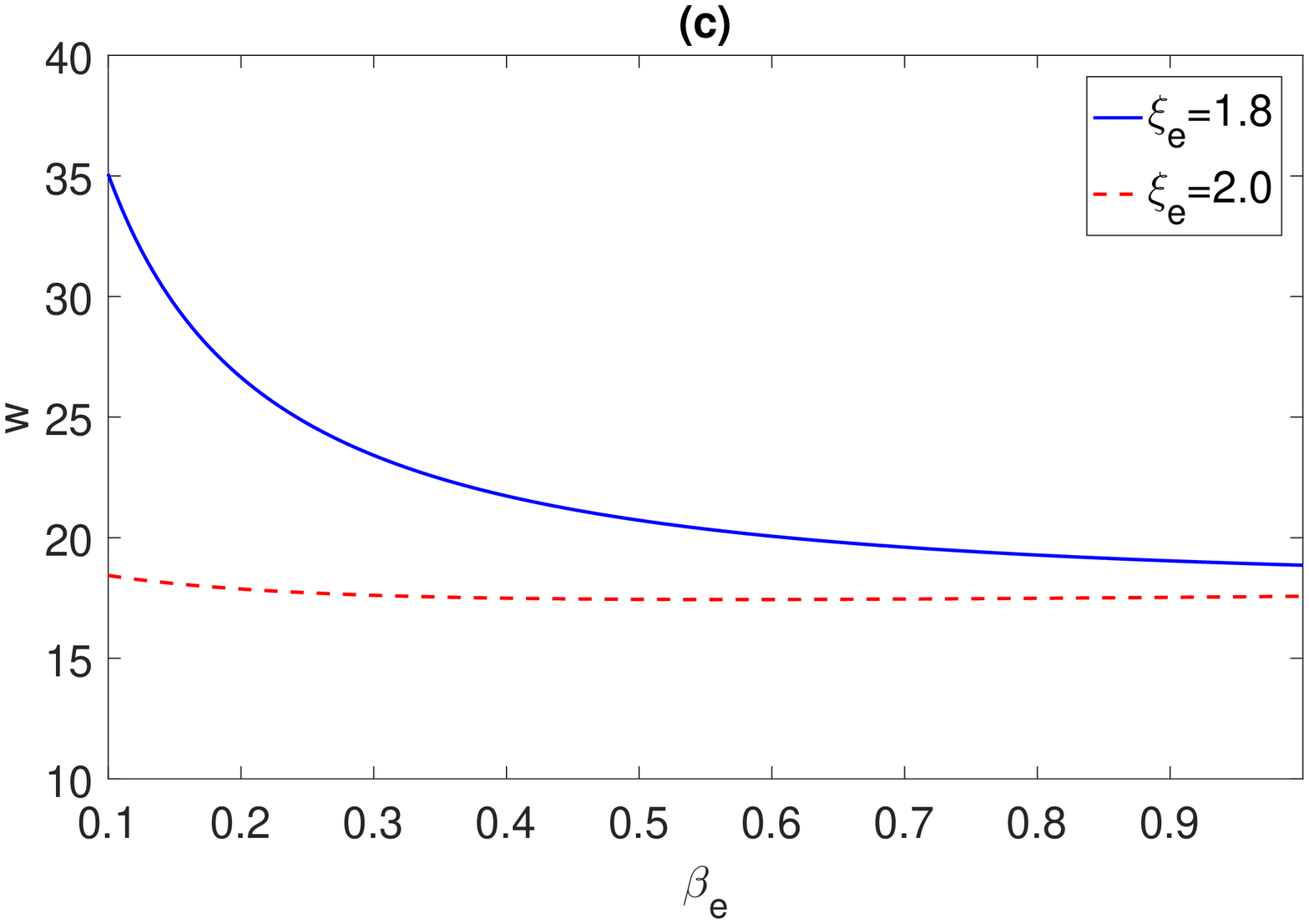}
\includegraphics[width=3.2in,height=2.0in]{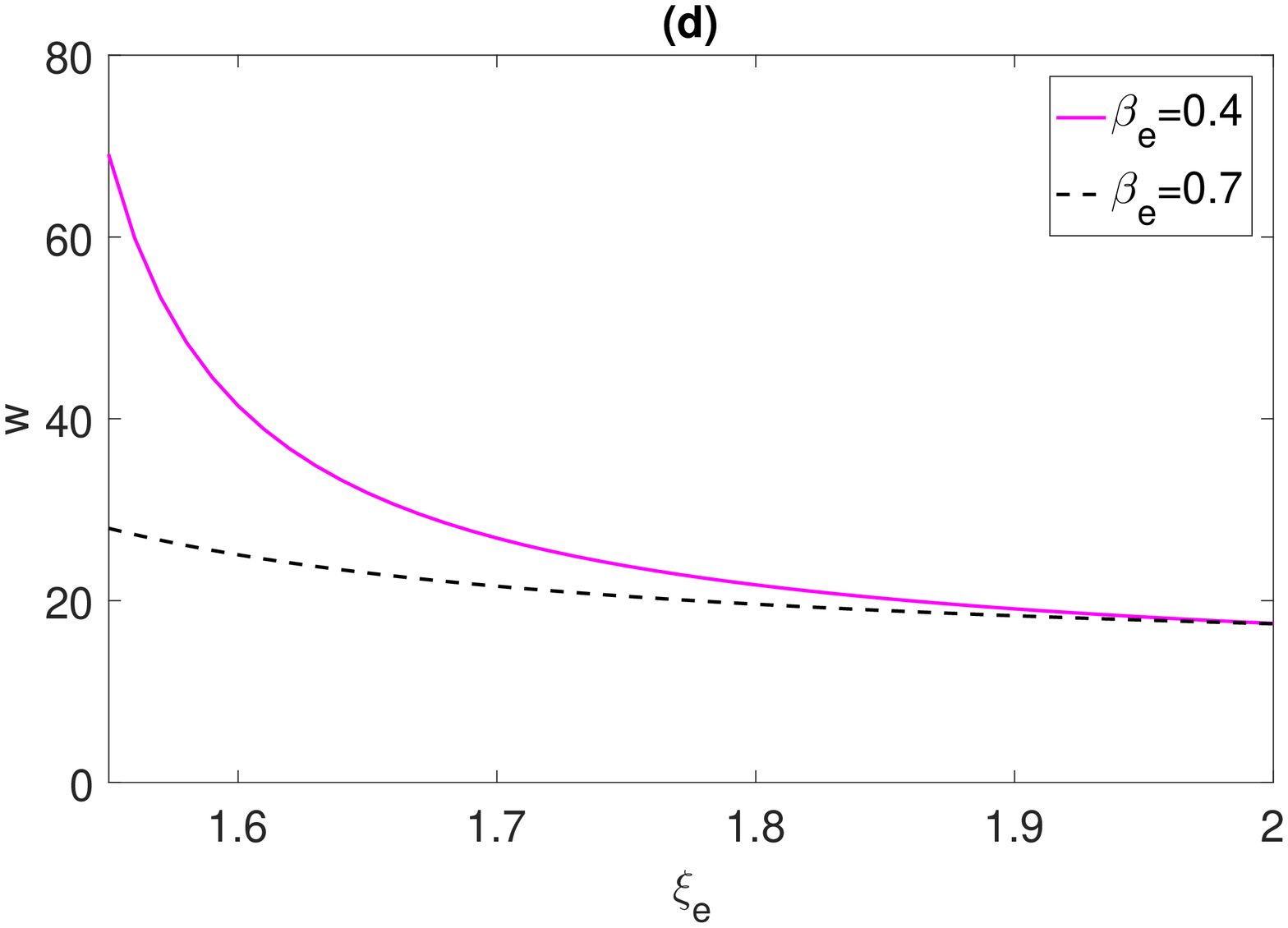}
\caption{The variation of the amplitude [subplots (a) and (b)] and the width [subplots (c) and (d)] of the KdV soliton [Eq. \eqref{eq-sol-kdv}] are shown for different values of $\beta_e$ ($<1$) and $\xi_e$ as in the legends. The fixed parameter values are $\delta=0.7$, $\sigma_i=0.5$,   $\sigma_p=0.8$, and $U=0.01$.}
\label{fig:amp_width_betae_xie}
\end{figure}

 
In what follows, we  study the influence of the relativity and degeneracy parameters $\beta_e$ and $\xi_e$ on the profiles of the soliton   amplitude  and width for a set of fixed parameter values, namely  $\sigma_i=0.5$,   $\sigma_p=0.8$, $\delta=0.7$, and $U=0.01$. We also   examine the  critical values of $\beta_e$ and $\xi_e$ below or above which the polarity of solitons   changes.  The results are displayed in Fig. \ref{fig:amp_width_betae_xie}. The subplots (a) and (b) show  the amplitudes $\phi_m$ and the subplots (c) and (d)   the widths $w$.  From subplots (a) and (b), it is seen that keeping any  one of $\xi_e$ and $\beta_e$ fixed and varying the other, there exists a critical value of $\xi_e$ or $\beta_e$ below (or above) which the raraefactive (or compressive) ion-acoustic solitons exist. Such a critical value of $\xi_e$ (or $\beta_e$) is upshifted    even  with a small reduction of $\beta_e$ (or $\xi_e$). Furthermore,   close to (or at) the critical values of $\beta_e$ and $\xi_e$ [e.g., for $\beta_{\rm{ec}}\sim0.34$, corresponding to the dashed line of subplot (a) and $\xi_{\rm{ec}}=1.97$, corresponding to the solid line in subplot (b)], a significant increase in magnitude of the soliton amplitude is seen. Also,  some subintervals of   $\beta_e$ and $\xi_e$ exist in each of which  the   amplitudes for rarefactive solitons are close to zero and so is the soliton energy. Such intervals corresponding to the solid lines in subplots (a) and (b), respectively, are
$0<\beta_e\lesssim0.3$ and $0<\xi_e\lesssim1.7$. In contrast, the amplitude of compressive solitons initially decreases but reaches a steady state as $\beta_e$ approaches the unity.   Thus, when the degeneracy parameter is fixed at $\xi_e=1.8$ (and other parameters as above) and the relativity parameter $\beta_e$ varies in $0<\beta_e\lesssim1$, the rarefactive solitons exist in $0.3\lesssim\beta_e\lesssim0.6$ and the compressive solitons exist in $0.7\lesssim\beta_e\lesssim1$ [See the solid line in subplot (a)].   \textit{On the other hand, if the relativity parameter is fixed at $\beta_e=0.7$ and other parameters as above, the rarefactive solitons exist in $1.5\lesssim\xi_e\lesssim1.7$ and the compressive solitons exist in $1.8\lesssim\xi_e\lesssim3$ [See the dashed line in subplot (b)]. Similar domains can be obtained for some other fixed values of $\xi_e$ and $\beta_e$.}
These parameter regimes are in agreement with our previous prediction (\textit{cf}. Fig.\ref{fig:A1_zero}).   From subplot (a), it can be inferred that  if the value of $\xi_e$ is further reduced from $\xi_e=1.8$ to $\xi_e=1.6$ (keeping the other parameters fixed as above or as in the figure caption), only the rarefactive solitons exist in  $0<\beta_e<1$.  However, no such domain of $\xi_e$ can be found from subplot (b) for which only the  rarefactive or compressive solitons can  exist. It is also evident from subplots (a) and (b) that while the  amplitude of compressive solitons decreases   and reaches a steady state with increasing values of either $\xi_e$ or $\beta_e$ above their critical values, the same (in magnitude) for the rarefactive solitons increases with increasing values of either $\xi_e$ or $\beta_e$ below their critical values.   These are also in agreement with  what we have predicted before from Fig. \ref{fig:A1_zero}. 
On the other hand, subplots (c) and (d) of Fig. \ref{fig:amp_width_betae_xie} show the variations of the soliton width with respect to the parameters $\beta_e$ and $\xi_e$. In both the cases it is seen that the soliton width decreases but reaches a steady state with increasing values of both $\beta_e$ and $\xi_e$.  
\par 
From the characteristics of the soliton amplitude and width as   in Fig. \ref{fig:amp_width_betae_xie}, it may be assessed that given  a   set of fixed parameter values  of  $\sigma_i$, $\sigma_p$, $\delta$, and $U$, there exist two subintervals   of both $\beta_e$ and $\xi_e$, namely $0<\beta_e<\beta_{\rm{ec}}$, $\beta_{ec}<\beta_e<1$ and   $0<\xi_e<\xi_{\rm{ec}}$, $\xi_{\rm{ec}}<\xi_e<3$. In $0<\beta_e<\beta_{\rm{ec}}$ and $0<\xi_e<\xi_{\rm{ec}}$, the energy of rarefactive solitons is initially very low. However, it starts increasing with increasing values of $\beta_{e}$ and $\xi_{e}$ below their critical values $\beta_{\rm{ec}}$ and $\xi_{\rm{ec}}$. Such solitons with growing amplitude and hence with increasing energy can evolve in relativistic degenerate plasmas but may be unstable. On the other hand, the    energy of compressive solitons  (although initially high) tends to decrease   in $\beta_{\rm{ec}}<\beta_e<1$ and $\xi_{\rm{ec}}<\xi_e<3$,  but reaches a steady state at higher values of $\xi_e$ and $\beta_e<1$. Such solitons can evolve with a permanent profile for a longer time in the parameter space and thus can be stable.  Thus,   in astrophysical environments where both the chemical energy and the thermal energy of electrons and positrons are close to their rest mass energy,  the ion-acoustic compressive solitons may be stable, while rarefactive solitons may become unstable with their increasing amplitudes.  
%%%%%%%%%%%%%%%%%%%%%%%
%%%%%%%%%%%%%%%%%%%%%
%%%%%%%%%%%%%%%%%%
%%%%%%%%%%%%%%%%%%%%%%%%%%%%%%%%%%%%%%
\subsubsection{mKdV solitons} \label{sec-mKdV}
In Sec. \ref{sec-kdv}, we have noted that the KdV equation is not valid for parameter values close to or at the critical points ($P=P_c$) on the curve $A_1 = 0$. In this situation one needs to deal with a different set of stretched coordinates and a different expansion scheme for the  evolution of small amplitude ion-acoustic perturbations.  In this section, we, however,  consider the case when the parameter values exactly lie on the line    $A_1 = 0$ and study the properties of ion-acoustic solitons at these critical parameter values.  The other case (``close to the critical points") will be studied in Sec. \ref{sec-gardner}.  To derive the mKdV equation for the evolution of ion-acoustic solitons at the critical points, we modify the stretched coordinates [corresponding to the higher order smallness of $k$, i.e., $k\sim{\cal O}(\epsilon)$] as 
\begin{equation}\label{eq-stretch-mkdv}
\xi=\epsilon(x-\lambda t),~\tau=\epsilon^3.
\end{equation}
However, we retain  the same perturbation expansion scheme for the dependent variables  and follow the same reductive perturbation technique as for the KdV equation \eqref{eq-kdv}. So, in the lowest order of $\epsilon$, we  obtain the  same expressions for $n_{j1}$ $(j=e,~p,~i)$, $v_{i1}$ and $\lambda$, as given in Eqs. \eqref{eq-1st-order} and \eqref{eq-lambda}.
From the next  order  of $\epsilon$, the second order perturbed quantities yield
\begin{equation}\label{eq-ni2-mkdv}
n_{i2}=\frac{(3\lambda^2\nu_e-\sigma_i)}{2(\lambda^2\nu_e-\sigma_i)^3}\phi_1^2+\frac{1}{(\lambda^2\nu_e-\sigma_i)}\phi_2,
\end{equation} 
\begin{equation}\label{eq-vi2-mkdv}
v_{i2}=\frac{\lambda(\lambda^2\nu_e+\sigma_i)}{2(\lambda^2\nu_e-\sigma_i)^3}\phi_1^2+\frac{\lambda}{(\lambda^2\nu_e-\sigma_i)}\phi_2,
\end{equation}
\begin{equation}\label{eq-phi2-mkdv}
\nu_e\left(\alpha_en_{e2}-\alpha_pn_{p2}-n_{i2}\right)=-\frac{1}{2}A_1\phi_1^2=0.
\end{equation} 
For the third order  perturbed quantities, we obtain 
\begin{equation}\label{eq-ni3-mkdv}
-\lambda\frac{\partial n_{i3}}{\partial \xi }+ \frac{\partial n_{i1}}{\partial \tau }+\frac{\partial}{\partial \xi}(v_{i1}n_{i2}+n_{i1}v_{i2}+v_{i3})=0,
\end{equation}
\begin{equation}\label{eq-vi3-mkdv}
\begin{split}
-\lambda\frac{\partial v_{i3}}{\partial \xi }+ \frac{\partial v_{i1}}{\partial \tau }+\frac{\partial }{\partial \xi}(v_{i1}v_{i2})+\frac{1}{\nu_e}\frac{\partial \phi_3}{\partial \xi}+\frac{\sigma_i}{\nu_e}\left(\frac{\partial n_{i3}}{\partial \xi }+\frac{\partial }{\partial \xi }(\frac{n_{i1}^3}{3}-n_{i1}n_{i2})\right)=0,
\end{split}
\end{equation}
\begin{equation}\label{eq-ne3-mkdv}
n_{e3}=a_{0e}\phi_3+2a_{1e}\phi_1\phi_2+a_{2e}\phi_1^3,
\end{equation}
\begin{equation}\label{eq-np3-mkdv}
n_{p3}=-\frac{a_{0p}}{\sigma_p}\phi_3+\frac{a_{1p}}{\sigma_p^2}(2\phi_1\phi_2)-\frac{a_{2p}}{\sigma_p^3}\phi_1^3,
\end{equation}
\begin{equation}\label{eq-phi3-mkdv}
\frac{\partial^2 \phi_1}{\partial \xi^2 }=\nu_e\left(\alpha_en_{e3}-\alpha_pn_{p3}-n_{i3}\right),
\end{equation}
where 
\begin{equation}\label{eq-a2j}
\begin{aligned}
a_{2j}=A_j \left[\frac{\xi_j \beta_j}{2}\left(\frac{5}{16}\xi_j^{-3}+\frac{5\pi^2}{128}\xi_j^{-5}+\frac{245\pi^4}{6144}\xi_j^{-7}\right)
-\left(\frac{1}{16}\xi_j^{-3}+\frac{5\pi^2}{128}\xi_j^{-5}+\frac{147\pi^4}{2048}\xi_j^{-7}\right) \right].
\end{aligned}
\end{equation}
Finally, eliminating the third-order quantities from Eqs. \eqref{eq-ni3-mkdv}-\eqref{eq-phi3-mkdv} by using Eqs. \eqref{eq-ni2-mkdv}-\eqref{eq-phi2-mkdv},  we obtain the following mKdV equation for the evolution of ion-acoustic solitons at the critical points $P=P_c$.
\begin{equation}\label{eq-mkdv}
\frac{\partial \psi}{\partial \tau }+A_2B\psi^2\frac{\partial \psi}{\partial \xi }+B\frac{\partial^3 \psi}{\partial \xi^3 }=0,
\end{equation} 
where $\psi\equiv\phi_1$ and 
\begin{equation}
A_2=3\nu_e\left[\frac{11\lambda^4\nu_e^2+(2\lambda^2\nu_e-\sigma_i)^2}{6(\lambda^2\nu_e-\sigma_i)^5}-\frac{\alpha_p a_{2p}}{\sigma_p^3}-\alpha_ea_{2e}\right].
\end{equation}
From Eq. \eqref{eq-mkdv}, we note that, not only the nonlinear coefficient of the mKdV equation is modified  to $A_2B$, but the nonlinearity is also of higher order (compared to that of the KdV equation) of the first order electrostatic perturbation. This is expected as we have redefined the new space and time scales slower than the previous ones [Eq. \eqref{eq-stretch}] and accordingly the nonlinear effects appear in the higher order of perturbations. 
\par
\begin{figure} 
\centering
\includegraphics[width=4.5in,height=2.5in]{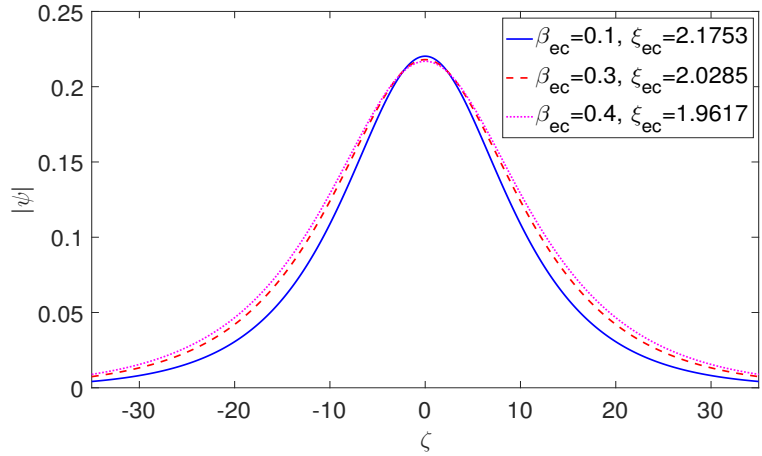}
\caption{The   profiles of the mKdV soliton [Eq. \eqref{eq-sol-mkdv}] are shown at different critical points (lying on the line $A_1=0$) as in the legend. The  fixed parameter values of $\delta$, $\sigma_i$, and $\sigma_p$ are as in Fig. \ref{fig:amp_width_betae_xie}.}
\label{fig:mKdV_SW_sigma_delta}
\end{figure}

   
In what follows, a stationary soliton solution of Eq. \eqref{eq-mkdv} (different from the KdV soliton) is given by 
\begin{equation}\label{eq-sol-mkdv}
\psi= \psi_m \rm{sech}\left(\zeta/w\right),
\end{equation} 
where $\psi_m$ $(= \pm \sqrt{6 U /A_2B})$ and $w$ $(=\sqrt{B/U})$ are, respectively, the amplitude and width of the  ion-acoustic mKdV soliton. We note that since $B$ is always positive, for real soliton solution, $A_2$ must be positive. Also, since $\psi_m$ can be both  positive and negative,   the coexistence of both the compressive and rarefactive solitons is possible at the critical points $P_c \equiv (\beta_{\rm{ec}}, \xi_{\rm{ec}})$ having the same  amplitude (in magnitude)   and the same width.  Furthermore, if at some  points $P$ (other than those satisfy $A_1=0$),  $A_2\rightarrow 0$, then $\psi_m \rightarrow \pm \infty$, implying that  the  mKdV equation  \eqref{eq-mkdv} may no longer be valid. So, in a situation when $A_1 \approx0$ and $ A_2 \approx0$, no finite soliton solution will exist and one thus has to   look for another evolution equation with further higher order corrections. However, this is not of interest to the present study.  In fact, we find that for a wide range of critical values  of the parameters (at which $A_1=0$), $A_2$ remains positive and finite  (See Table \ref{table:1}). The latter also ensures the existence of both compressive and rarefactive ion-acoustic mKdV solitons.      The typical profiles of the  mKdV soliton at different critical points $P_c$ are shown in Fig.  \ref{fig:mKdV_SW_sigma_delta}. Since the polarity of the soliton changes only due to the sign change in $\psi_m$, we plot only the profiles   for $|\psi|$ against $\zeta$. From Fig.  \ref{fig:mKdV_SW_sigma_delta} it is found that although the  amplitude can be slightly modified, the mKdV solitons can be wider at a critical point with a higher value of $\beta_{\rm{ec}}$, but a lower value of $\xi_{\rm{ec}}$ as in the legend. At these critical values, the soliton energy will also be higher. However, since the  amplitude does not change significantly, the soliton can evolve    with a stable profile. We recall that for values of the parameters near the critical points $P_c$ of $A_1=0$, the KdV and mKdV equations do  not give any finite soliton solution. In such a situation we need to look for another  evolution equation, namely the Gardner equation which we will derive in Sec. \ref{sec-gardner}.   
%%%%%%%%%%%%%%%%%%%% 
\begin{table}[ht]
\centering
\begin{tabular}{c c c c c c } 
\hline
 $\sigma_i$ \hspace{0.3cm} & $\sigma_p$ \hspace{0.3cm} &  $\delta$  \hspace{0.3cm} &  $\xi_e$  \hspace{0.3cm} &  $\beta_e$ \hspace{0.3cm} & $A_2$   \\
\hline
0.5 \hspace{0.3cm} & 1.5 \hspace{0.3cm} & 0.6 \hspace{0.3cm} &\multirow{1}{1.2cm}{1.7628}  \hspace{0.3cm}&  0.2 &3.5804  \\
& & & 1.6465 \hspace{0.3cm} & 0.3 &4.2117  \\

0.5 \hspace{0.3cm} & 0.8 \hspace{0.3cm} & 0.6 \hspace{0.3cm} &\multirow{1}{1.2cm}{2.0641}  \hspace{0.3cm}&  0.1 &1.9423  \\
& & & 1.9765 \hspace{0.3cm} & 0.2 &1.8049  \\

0.5 \hspace{0.3cm} & 0.8 \hspace{0.3cm} & 0.7 \hspace{0.3cm} &\multirow{1}{1.2cm}{2.0285}  \hspace{0.3cm}&  0.3 &1.7112  \\
& & & 1.9617 \hspace{0.3cm} & 0.4 &1.5706  \\

0.9 \hspace{0.3cm} & 0.8 \hspace{0.3cm} & 0.7 \hspace{0.3cm} &\multirow{1}{1.2cm}{2.0736}  \hspace{0.3cm}&  0.2 &2.1609  \\
& & & 1.8124 \hspace{0.3cm} & 0.6 &1.5795  \\
\hline

\end{tabular}

\caption{The values of the nonlinear coefficient $A_2 $ of the mKdV equation are presented against different parametric values.}
 \label{table:1} 
\end{table}
%%%%%%%%%%%%%%%%%%%%%%
%%%%%%%%%%%%%%%%%%%%%%%%%%%%%%%%%%%%%%   

\subsubsection{Gardner solitons} \label{sec-gardner}
In this section, we study the evolution of ion-acoustic solitons in the parameter space of  $A_1\approx 0$. In the latter, both the KdV amd mKdV equations fail to describe the evolution of ion-acoustic solitons.   So, in order to explore the finite amplitude solitons beyond the KdV and mKdV limits, we derive the standard Gardner equation. To this end, we assume that  around the critical points  $P=P_c$ of $A_1=0$,  $A_1 \simeq s \epsilon$, where  $s=1$ for $A_1>0$ and $s=-1$ for $A_1<0$. Since $A_1$ appears only in the perturbation equation of the Poisson equation, [\textit{cf}.   Eq. \eqref{eq-phi2-mkdv}], the second order perturbed quantities give  
\begin{equation}\label{eq-phi2-gardner}
\epsilon^2\nu_e\left(\alpha_en_{e2}-\alpha_pn_{p2}-n_{i2}\right)\approx\frac{1}{2}\epsilon^3 s\phi_1^2,
\end{equation} 
i.e., they   appear in the third order of $\epsilon$.  This should be included in the third order correction equation of the Poisson equation. 
%\begin{equation}\label{approx of total number density}
%\epsilon^2\rho_2\approx\frac{1}{2}\epsilon^3   s\psi^2,
%\end{equation}
%which must be included in the third order Poisson equation.
As before,   the first order quantities will remain the same as for the KdV and mKdV equations. Also, the second order perturbations will give the same results as for the mKdV equation. So, for the third order perturbed quantities, we obtain from  Eqs. \eqref{eq-e-moment}-\eqref{eq-poisson-dimless} and \eqref{eq-nj2} the following: 
 \begin{equation}\label{eq-ni3-gardner}
-\lambda\frac{\partial n_{i3}}{\partial \xi }+ \frac{\partial n_{i1}}{\partial \tau }+\frac{\partial}{\partial \xi}(v_{i1}n_{i2}+n_{i1}v_{i2}+v_{i3})=0,
\end{equation}
\begin{equation}\label{eq-vi3-gardner}
\begin{split}
-\lambda\frac{\partial v_{i3}}{\partial \xi }+ \frac{\partial v_{i1}}{\partial \tau }+\frac{\partial }{\partial \xi}(v_{i1}v_{i2})+\frac{1}{\nu_e}\frac{\partial \phi_3}{\partial \xi}+\frac{\sigma_i}{\nu_e}\left(\frac{\partial n_{i3}}{\partial \xi }+\frac{\partial }{\partial \xi }(\frac{n_{i1}^3}{3}-n_{i1}n_{i2})\right)=0,
\end{split}
\end{equation}
\begin{equation}\label{eq-poiss-gardner}
\frac{\partial^2 \phi_1}{\partial \xi^2 }=\nu_e\left(\alpha_en_{e3}-\alpha_pn_{p3}-n_{i3}-\frac{1}{2}\frac{s}{\nu_e}\phi_1^2\right),
\end{equation}
\begin{equation}\label{eq-ne3-gardner}
n_{e3}=a_{0e}\phi_3+2a_{1e}\phi_1\phi_2+a_{2e}\phi_1^3,
\end{equation}
\begin{equation}\label{eq-np3-gardner}
n_{p3}=-\frac{a_{0p}}{\sigma_p}\phi_3+\frac{a_{1p}}{\sigma_p^2}(2\psi\phi_2)-\frac{a_{2p}}{\sigma_p^3}\phi_1^3.
\end{equation}
Finally, eliminating the third order perturbations by using the second order perturbed quantities as in Sec. \ref{sec-mKdV}, from Eqs. \eqref{eq-ni3-gardner}-\eqref{eq-np3-gardner} we obtain   the following   Gardner equation.
\begin{equation}\label{eq-gardner}
\frac{\partial \psi}{\partial \tau }+s   B\psi\frac{\partial \psi}{\partial \xi }+A_2B\psi^2\frac{\partial \psi}{\partial \xi }+B\frac{\partial^3 \psi}{\partial \xi^3 }=0,
\end{equation} 
where, as before, $\psi=\phi_1$. 
\par 
From Eq. \eqref{eq-gardner}, we note that  in comparison with the mKdV equation \eqref{eq-mkdv}, an additional nonlinearity proportional to $s$ [similar to the KdV equation \eqref{eq-kdv}]  appears. Accordingly, Eq. \eqref{eq-gardner}   is often called the KdV-mKdV equation.  The  additional term (proportional to $s$) appears in Eq. \eqref{eq-gardner} due to the smallness of $A_1$:  $A_1\sim{\cal O}(\epsilon)$, i.e.,  $A_1\neq 0$.     So, the  Gardner equation is valid for the parametric values close to the critical points of the curve  $A_1=0$ (Fig.  \ref{fig:A1_zero}). In particular, for $A_2 \rightarrow 0$, the Gardner equation reduces to the KdV equation \eqref{eq-kdv} with the nonlinear  coefficient $A_1 \nsim {\cal O}(\epsilon)$ and with the  same solution [Eq. \eqref{eq-sol-kdv}] for the  finite amplitude ion-acoustic solitons in relativistic degenerate plasmas. 
\par
\begin{figure}! 
\centering
\includegraphics[width=3.2in,height=2.5in]{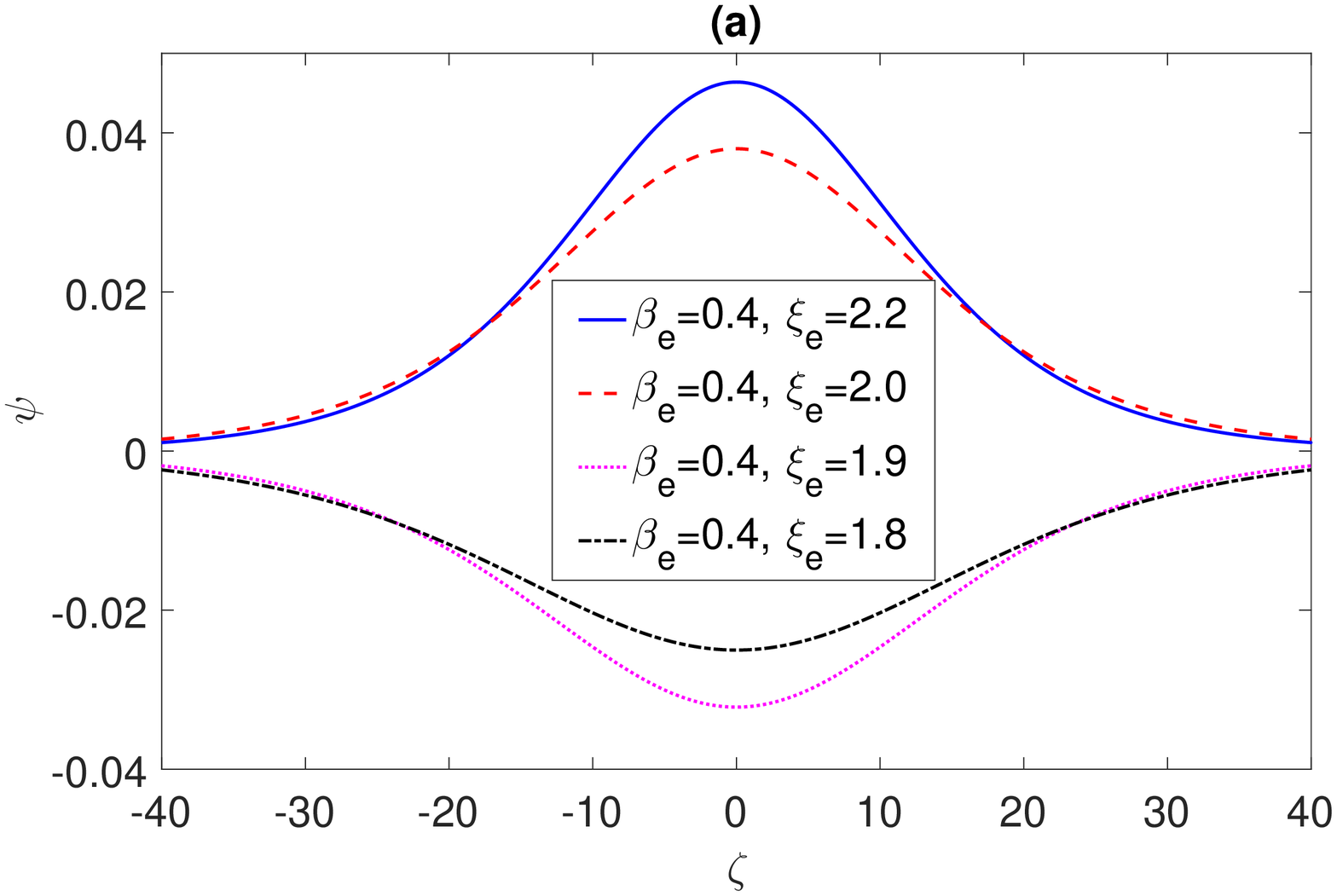}
\includegraphics[width=3.2in,height=2.5in]{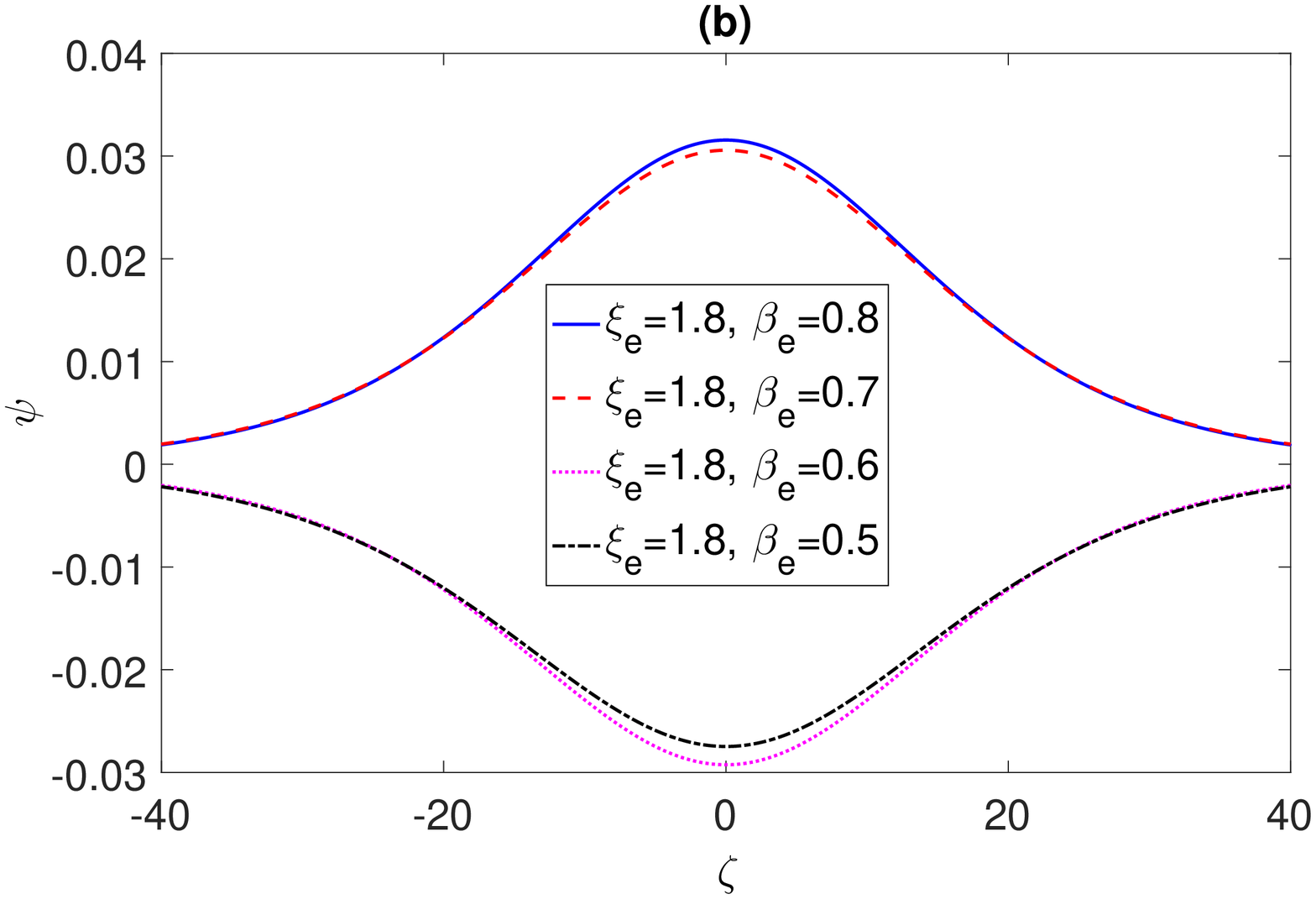}
\caption{The  profiles of the Gardner solitons [Eq. \ref{eq-sol-gardner}] are shown  at different points $(\beta_e,\xi_e)$ that are close to the critical points, i.e., when   $A_1\sim {\cal O}(\epsilon)$. The fixed parameter values are  $\delta=0.7$, $\sigma_i=0.5$, and $\sigma_p=0.8$. }
\label{fig:gardner_SW}
\end{figure}

 
A  stationary soliton solution of Eq. \eqref{eq-gardner} can also be obtained by  using the transformation $\zeta=\xi-U \tau$     as follows. Under this transformation, Eq. \eqref{eq-gardner} reduces to
\begin{equation}\label{eq-energ-int}
\frac{1}{2}\left(\frac{d\psi}{d\zeta}\right)^2+V(\psi)=0,
\end{equation}  
 where the pseudo-potential $V(\psi)$  (with $U>0$ and $B>0$) is given by  
\begin{equation}\label{eq-V}
V(\psi)=-\frac{U}{2B}\psi^2+\frac{s}{6}\psi^3+\frac{A_2}{12} \psi^4.
\end{equation} 
  For the existence of soliton solutions, it is necessary for $V(\psi)$ to satisfy the following conditions for some $\psi=\psi_m$ $\neq 0$. 
\begin{equation}\label{eq-V-cond1}
V(\psi)|_{\psi=0}=\frac{dV(\psi)}{d\psi}|_{\psi=0}=0,
\end{equation}
\begin{equation}\label{eq-V-cond2}
\frac{d^2V(\psi)}{d\psi^2}|_{\psi=0}<0,
\end{equation}
\begin{equation}\label{eq-V-cond3}
V(\psi)|_{\psi=\psi_m}=0.
\end{equation}
It is straightforward to show that the first two conditions are eventually satisfied.  The third condition gives
\begin{equation}\label{eq-psim12}
\psi_{{m_1},{m_2}}=\psi_0\left[1\mp\sqrt{1+U/V_0}\right],
\end{equation}
where $\psi_0=-s/A_2$ and $V_0=s^2B/{6A_2}$.
Thus,   Eq. \eqref{eq-energ-int} reduces to
\begin{equation}\label{eq-psi-gardner}
\left(\frac{d\psi}{d\xi}\right)^2+r \psi^2(\psi-\psi_{m1})(\psi-\psi_{m2})=0,
\end{equation}
where  $r=A_2/6$. The soliton solution of  Eq. \eqref{eq-psi-gardner} can then be obtained as
\begin{equation}\label{eq-sol-gardner}
\psi=\left[\frac{1}{\psi_{m_2}}-\left(\frac{1}{\psi_{m_2}}-\frac{1}{\psi_{m_1}}\right)\cosh^2\left(\frac{\zeta}{w}\right)\right]^{-1},
\end{equation}
where $w=2/\sqrt{-r \psi_{m_1}\psi_{m_2}}=2\sqrt{B/U}$.   The  profiles of the Gardner soliton [Eq. \eqref{eq-sol-gardner}]  are shown in Fig. \ref{fig:gardner_SW} at  different points $P\equiv(\beta_e,\xi_e)$ that are close to the critical point $P_c$ of the curve $A_1=0$.  We choose the fixed values as  $\sigma_i=0.5$,   $\sigma_p=0.8$, and $\delta=0.7$ and consider a pair of different sets of values of $\beta_e$ and $\xi_e$ for which both the compressive and rarefactive solitons can coexist. It is found that while the width remains almost unchanged, the amplitudes of both the compressive and rarefactive solitons increase (and hence   solitons can evolve with increasing energies) with increasing values of $\xi_e$ [subplot (a)] and $\beta_e$ [subplot (b)].  
%Since the parameter values are to be restricted in the region for which $A_1\sim{\cal O}(\epsilon)$
%%%%%%%%%%%%%%%%%%
%%%%%%%%%%%%%%%%%%%%%%%%%%
%%%%%%%%%%%%%%%%%%%%%%%%%%%%%%%%%%%%%%%%%%%%%%%%%%%%%%%%%
\subsection{Case II:   $\beta_e>1$} \label{sec-case2}
%%%%%%%%%%%%%%%%%%%%%%%%%%%%%%%%%%%%%%%%
\begin{figure} 
\centering
\includegraphics[width=4.5in,height=2.5in]{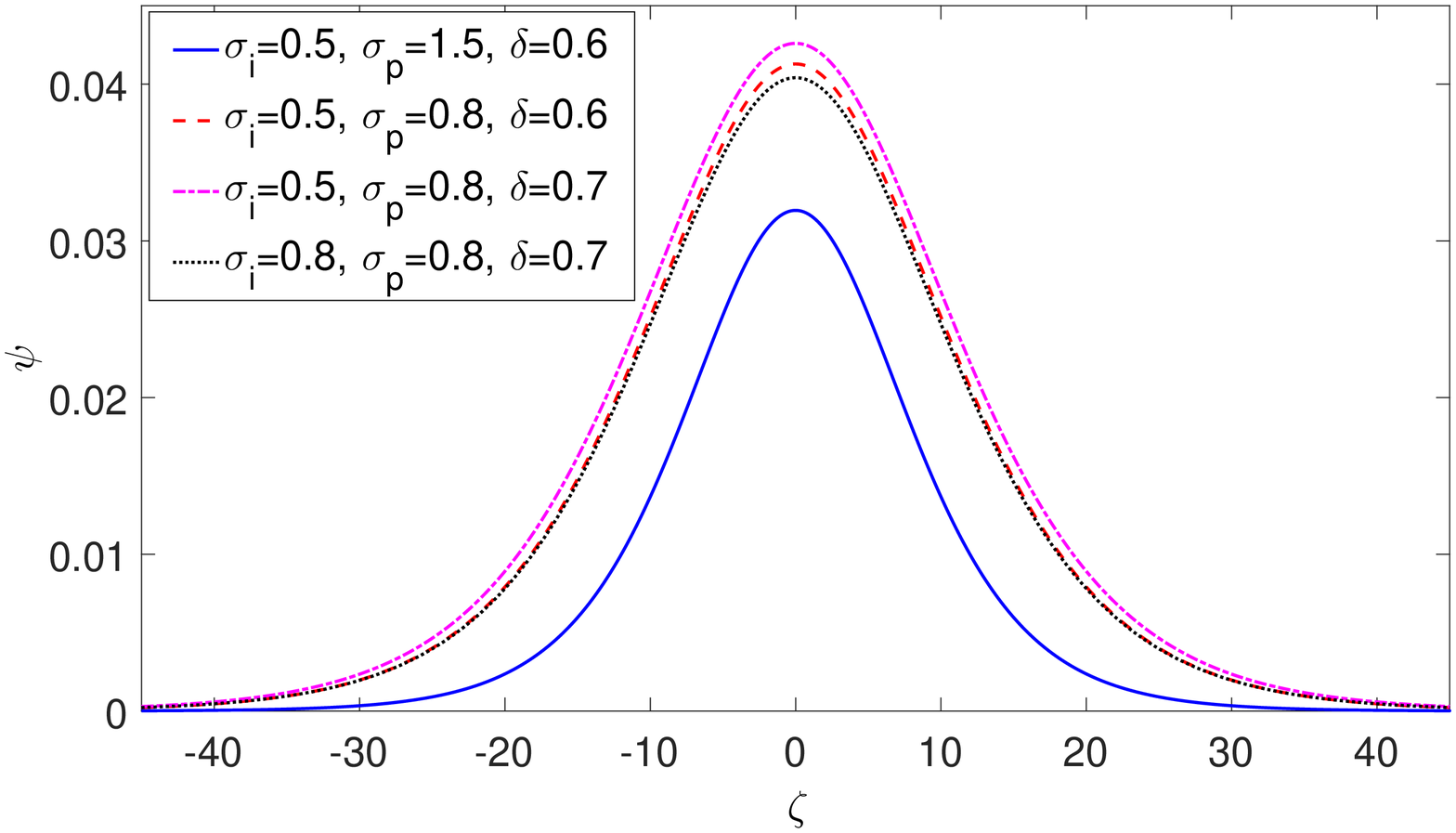}
\caption{The profiles of the compressive KdV solitons [The case of $\beta_e>1$] are shown for different values of $\sigma_i$,   $\sigma_p$, and $\delta$. The other parameter values are $U=0.01$, $\beta_e=2$, and $\xi_e=3$. }
\label{fig:KdV_SW_beta_gt_1}
\end{figure}
%%%%%%%%%%%%%%%%%%%%%%%%%%%%%%%%%%%%%%%%%%%%%%%%%%%%%%%%%%
\begin{figure} 
\centering
\includegraphics[width=3.2in,height=2.2in]{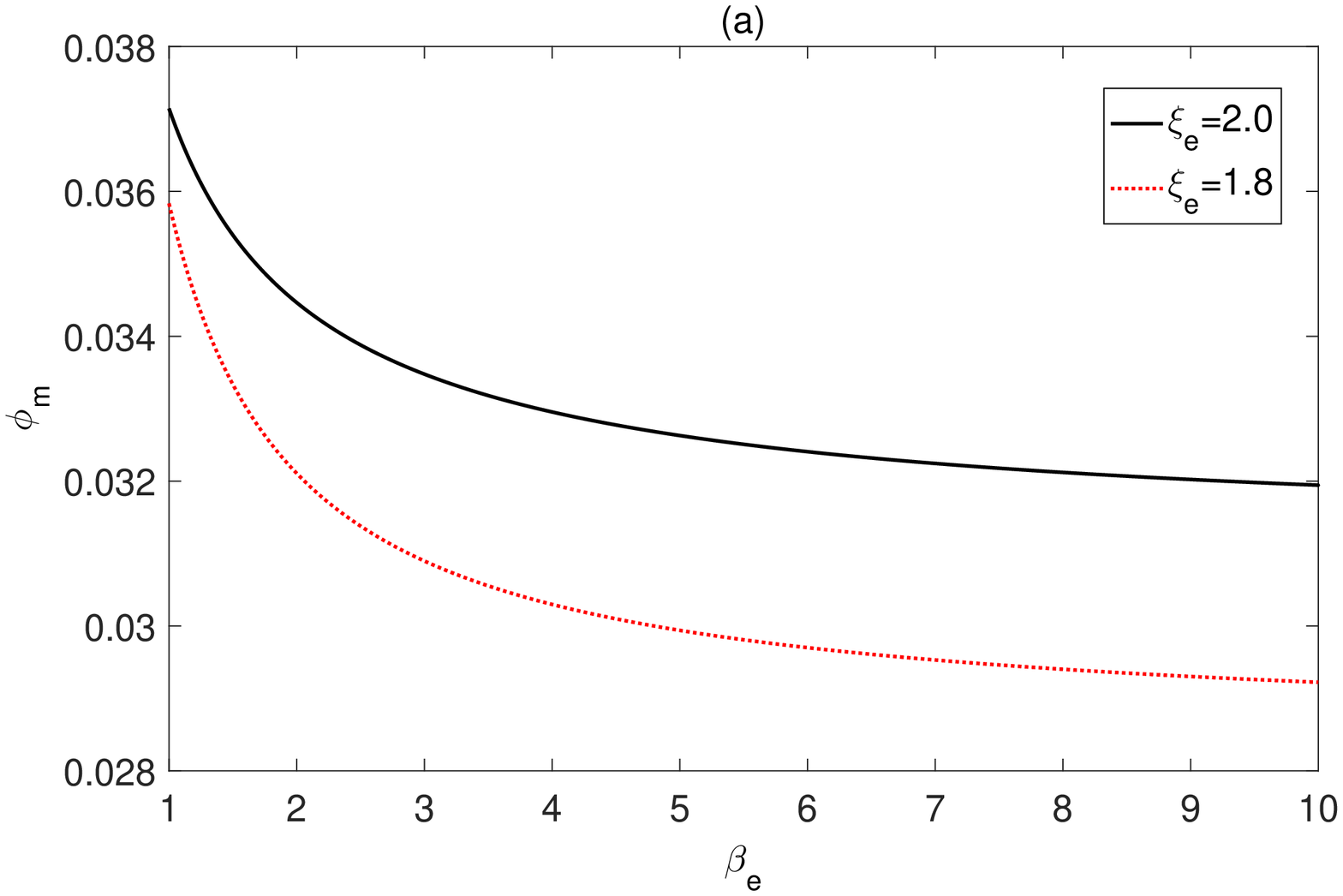}
\includegraphics[width=3.2in,height=2.2in]{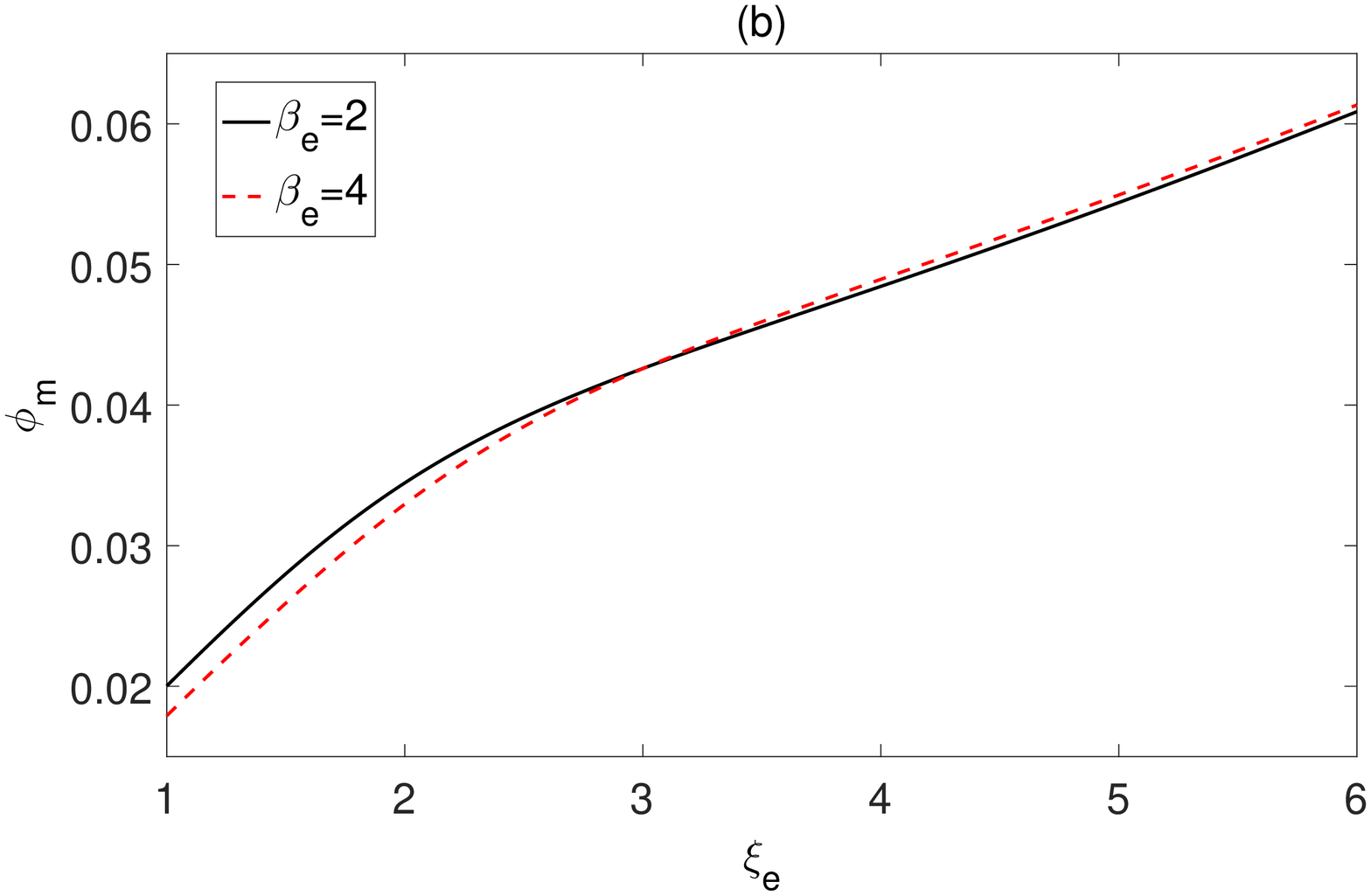}
\includegraphics[width=3.2in,height=2.2in]{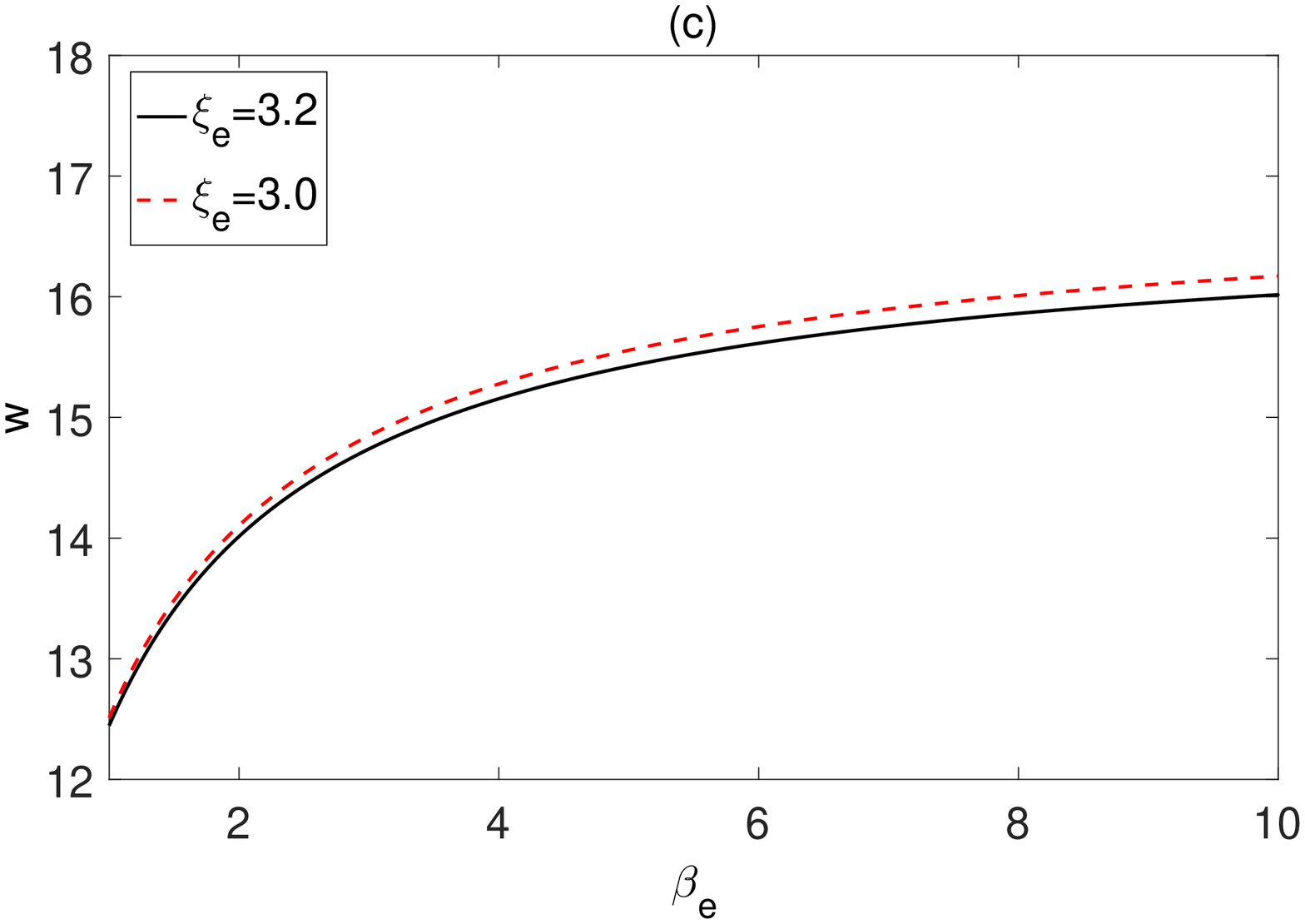}
\includegraphics[width=3.2in,height=2.2in]{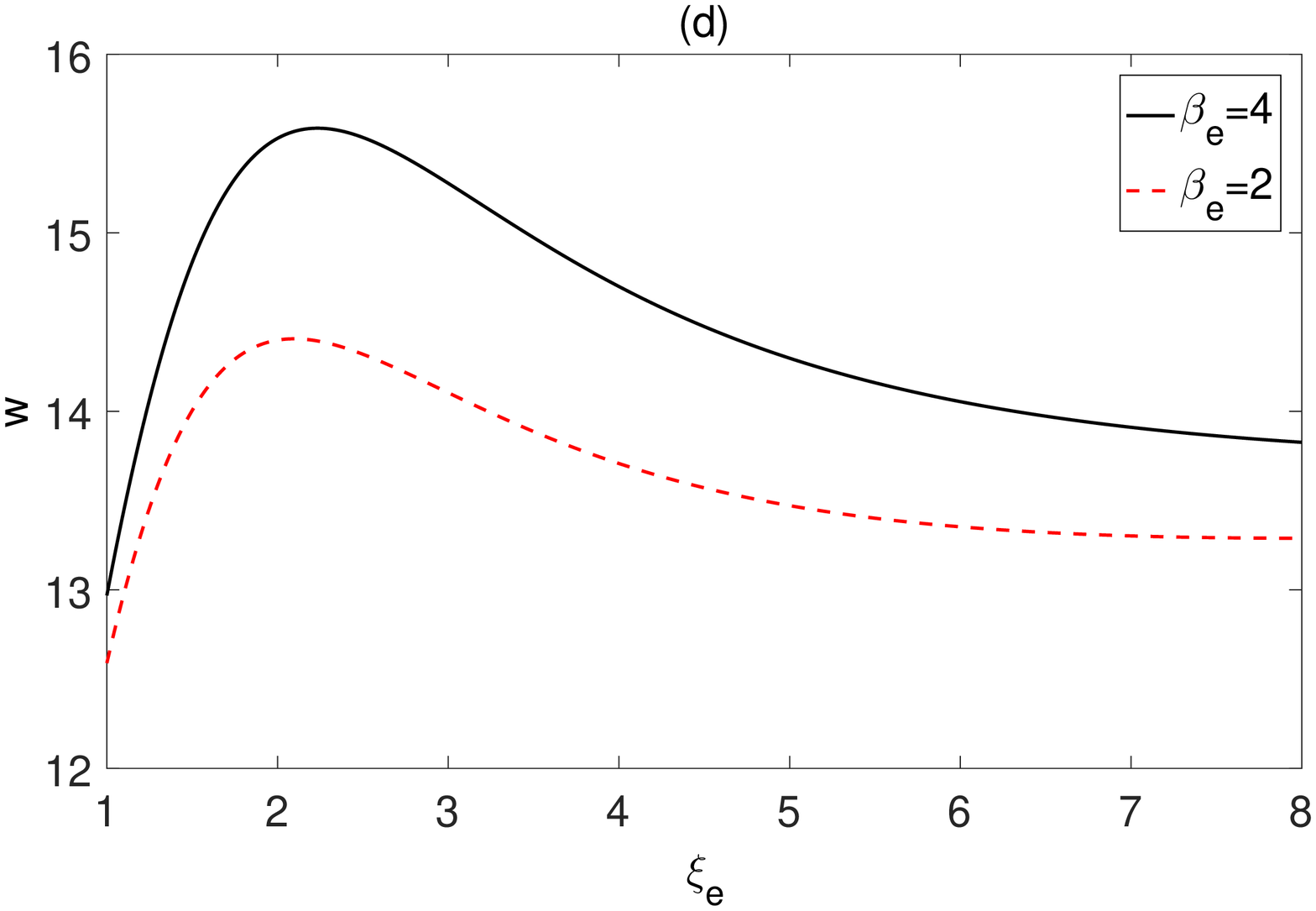}
\caption{The variation of the soliton amplitude [Subplots (a) and (b)] and width [Subplots (c) and (d)] of the KdV compressive soliton (The case of $\beta_e>1$) are shown for different values of $\beta_e$ ($>1$) and $\xi_e$ as in the legends. The fixed parameter values are $\delta=0.7$, $\sigma_i=0.5$, $\sigma_p=0.8$, and $U=0.01$.}
\label{fig:amp_width_betae_xie_gt_1}
\end{figure}
%%%%%%%%%%%%%%%%%%%%%%%%%%%%%%%%%%%%%%%%%%%%%%%%%%%%%%
We move to the Case II with $\beta_e>1$, i.e., when the thermal energy of electrons/positrons is slightly larger than their rest mass energy. Going back to the coefficients of the KdV equation \eqref{eq-kdv}, we find that apart from $B>0$, $A_1$ remains positive and finite for $\beta_e>1$. Consequently, the KdV equation \eqref{eq-kdv} remains valid in this case and so is its   solution \eqref{eq-sol-kdv} with $A_1>0$. It follows that the parameter regimes in Case II only support the existence of compressive  ion-acoustic solitons in relativistic degenerate plasmas at finite temperature.       The typical potential profiles of the compressive solitons for $\beta_e>1$  are shown in Fig. \ref{fig:KdV_SW_beta_gt_1} for different values of $\sigma_i$, $\sigma_p$ and $\delta$. It is found that similar to the case of $\beta_e<1$ [Fig. \ref{fig:KdV_SW_sigma_delta} (a)] both the amplitude and width of the soliton profiles decrease with an increasing value of each of the parameters $\sigma_i$ and $\sigma_p$. However, these amplitudes and widths are found to be increased with increasing values of $\delta$.  Similar characteristics with increasing amplitudes and widths of solitons were  observed  in  Maxwellian e-p-i plasmas \cite{popel1995ion}. To study, in more details, the influences of the relativistic and degeneracy parameters $\beta_e$ ($>1$) and $\xi_e$ on the profiles of the soliton amplitude and width, we plot $\phi_m$ and $w$,   given in Sec. \ref{sec-kdv}, against $\beta_e$ and $\xi_e$. The results are shown in Fig. \ref{fig:amp_width_betae_xie_gt_1}.  Here, we consider the other parameter values fixed at  $\sigma_i=0.5$,   $\sigma_p=0.8$, and $\delta=0.7$. It is found that the characteristics of the soliton amplitude and width significantly differ from those   in the case of $\beta_e<1$ [See Fig. \ref{fig:amp_width_betae_xie}]. From subplot (a), it is seen that although the amplitude grows initially, it reaches a steady state at higher values of $\beta_e$, which may be a requisite condition for solitons to be stable with their finite energy even in strongly relativistic regime with $\beta_e>1$. On the other hand, the amplitude can grow with increasing values of $\xi_e$ within the domain, which may eventually lead to an instability due to increasing soliton energy. From subplots (c) and (d), one can observe that even though the width increases initially, it approaches a constant value with higher values of both $\beta_e$ and $\xi_e$.  Thus, it may be concluded that although the soliton amplitude increases with $\xi_e$, the amplitude remains finite and small even at large $\xi_e\sim20$. The latter can  be achieved at small electron (or positron) thermal energy ($\sim0.1$ ev) \cite{shi2014}. So, even in the case of strong relativistic degeneracy, ion-acoustic solitons   having finite energy can be stable.  However, the detailed discussion about the stability of ion-acoustic solitons is beyond the scope of the present investigation.
%%%%%%%%%%%%%%%%%%%%%
\section{Summary and conclusion} \label{sec-summary-conclu}
We have investigated the linear and nonlinear properties of ion-acoustic solitary waves in a multi-component relativistic degenerate plasma consisting of relativistic inertialess unmagnetized degenerate electrons and positrons at finite temperature and nonrelativistic classical thermal ions. Specifically, we have focused on the intermediate regimes relevant for astrophysical plasmas, e.g., in the core of white dwarfs, where the particle Fermi energy and the thermal energy do  not differ significantly, i.e., $T_{Fj}>T_j$ and the particle thermal energy is also close to the rest mass energy, i.e., $\beta_j\equiv k_BT_j/mc^2\sim1$. Depending on whether the ratio   $\beta_j$ is smaller or larger than unity and the other parameter restrictions applicable for the validity of the fluid model,  we have mainly classified two parameter regimes, namely Case I and Case II as in Sec. \ref{sec-phys-regime}, that are relevant in astrophysical environments (e.g., in the core of white dwarf stars).  The existence of ion-acoustic linear wave modes as well as the nonlinear evolution of ion-acoustic solitons are then studied in these two cases. The main theoretical results, so obtained in the linear and nonlinear regimes, are summarized as follows:
\begin{itemize}
\item[(a)] In the linear theory, a general dispersion relation is derived for ion-acoustic waves, which is shown to be significantly modified by the relativity parameter $\beta_j$, the degeneracy parameter $\xi_j\equiv \mu_j/k_BT_j$, the positron to electron and the ion to electron temperature ratios $\sigma_p$ and $\sigma_i$, and the positron to electron number density ratio  $\delta$. 
\item[$\bullet$]  It is found that the effective charge screening length  is significantly reduced due to the presence of the positron species $(\delta)$ in plasmas when $\beta_e<1$ (Case I).    However, the same is enhanced for plasma parameters satisfying $\beta_e>1$ (Case II).
\item[$\bullet$] It is shown that the ion-acoustic wave frequency is increased (decreased) with a small increase of $\beta_e$ $(\xi_e)$. Also,  similar to  classical electron-ion plasmas,   the linear phase velocity of ion-acoustic waves  in relativistic degenerate plasmas at finite temperature lies in $v_{ti}<\omega/k<\sqrt{c_s^2+v_{ti}^2}$ and is well below the speed of light in vacuum when $\beta_e\sim1$. However, $c_s$ is significantly different from that in classical electron-ion plasmas and has different expressions in two different parameter regimes (Case I and Case II).
\item[$\bullet$] The influences of the parameters $\sigma_p$ and $\sigma_i$ on the ion-acoustic wave frequency are found to be small. However, the effect becomes significant in the regime of higher values of $k>1$, which may not be relevant  for low-frequency ion-acoustic waves with longer wavelengths.
\item[(b)] In the nonlinear regime, the standard reductive perturbation technique is employed to derive the evolution equation for small amplitude ion-acoustic solitons, namely the KdV equation. Special attention is given to the critical cases where the KdV equation fails, but mKdV and Gardner equations can describe the evolution of ion-acoustic solitons. Nevertheless, the evolution of ion-acoustic solitons  in  two different parameter regimes (Case I and Case II) is significantly different. 
\item[$\bullet$] It is found that while the KdV equation applies for the Case II, the evolution of ion-acoustic solitons  in   the other parameter regime  (Case I)   can not be described by the KdV equation only, especially when the nonlinear effects tend to vanish in some critical parameter regions. In the latter, the evolution dynamics is rather described by the mKdV and Gardner equations.
\item[(d)] Case I, KdV solitons: 
%\item[$\bullet$]
\item[$\bullet$] It is found that the KdV equation can admit both compressive (with positive potential) and rarefactive (with negative potential) solitons. In a specific region of $\beta_e\xi_e$ plane,    higher the concentration of the positron species $(\delta)$ or lower the ion to electron temperature ratio $(\sigma_i)$,   more likely is the existence of rarefactive ion-acoustic solitons than the compressive solitons. 
\item[$\bullet$] The soliton energy can be  significantly reduced when the  positron species  has higher concentration and   thermal energy close to that of electrons.   Furthermore, the relative influence of the plasma parameters $\sigma_p$ and $\delta$ on the profiles of the compressive and rarefactive solitons are significantly different.
\item[$\bullet$] It is observed that for a fixed value of $\xi_e$:  $\xi_e=1.8$ (and other parameters fixed at   $\sigma_i=0.5$, $\sigma_p=0.8$, $\delta=0.7$, and   
$U=0.01$),  the rarefactive solitons exist in the regime $0.3\lesssim\beta_e\lesssim0.6$ and the compressive solitons exist in $0.7\lesssim\beta_e\lesssim1$. On the other hand, if the relativity parameter is fixed at $\beta_e=0.7$ and other parameters as above, the rarefactive solitons exist in $1.5\lesssim\xi_e\lesssim1.7$ and the compressive solitons exist in $1.8\lesssim\xi_e\lesssim3$.
\item[$\bullet$] In a regime where both the chemical energy and the thermal energy of electrons and positrons are close to the  rest mass energy,  the energy of ion-acoustic compressive solitons reaches a steady state value, while that of rarefactive solitons is increased   with their increasing amplitudes. It follows that the ion-acoustic compressive solitons may be stable, while rarefactive solitons may become unstable with higher energies.

\item[(e)] Case I, mKdV solitons: 
\item[$\bullet$] The mKdV solitons  are  valid only at the critical parameter values $\xi_{\rm{ec}}$ and $\beta_{\rm{ec}}$ of $\xi_{e}$ and $\beta_{e}$ where the KdV equation fails to describe the evolution of ion-acoustic solitons.  It is seen that both the compressive and rarefactive solitons can exist having the same amplitudes and widths. 
\item[$\bullet$] It is also found that although the amplitude can be  modified, the mKdV solitons can be wider and can have higher energies at a critical point with a higher value of $\beta_{\rm{ec}}$,  but a lower value of $\xi_{\rm{ec}}$.   Since the amplitude does not change significantly, the solitons can evolve with a stable profile. 
\item[(f)] Case I, Gardner solitons: 
\item[$\bullet$] The Gardner solitons  are  valid only at the  parameter values close to the critical values $\xi_{\rm{ec}}$ and $\beta_{\rm{ec}}$  where the KdV equation  also fails to describe the evolution of ion-acoustic solitons. It is found the qualitative features of both the compressive and rarefactive solitons are the same. However,  while the width remains almost unchanged, the amplitudes of both the compressive and rarefactive solitons increase and hence an increase of the soliton energy with increasing values of $\xi_e$   and $\beta_e$.  
\item[(g)] Case II, KdV solitons: 
\item[$\bullet$] In contrast to the Case I, only the compressive solitons are found to exist.  It is seen that although the soliton amplitude increases with $\xi_e$, but it remains finite and small even at large $\xi_e\sim20$. The latter can  be achieved at a small electron (or positron) thermal energy ($\sim0.1$ ev) \cite{shi2014}. So, even in the regime of strong relativistic degeneracy, the ion-acoustic solitons   having finite energy can be stable.
\end{itemize} 
\par 
To conclude, the relativistic high-density degenerate plasmas deviating from the thermodynamic equilibrium can appear not only in the context of laser produced plasmas or beam driven plasmas, but also in compact astrophysical objects like white dwarf stars, neutron stars. Such plasmas can support the propagation of low-frequency ion-acoustic waves   and hence ion-acoustic solitons as localized bursts of different radiation spectra emanating from these compact objects. So, the present results should be useful for understanding the localization of ion-acoustic solitary waves in these astrophysical environments.  Since we have considered the intermediate regime where $\beta_j\sim1$, the expressions for the electron and positron number densities [Eqs. \eqref{eq-nj1}] may not be applicable for the extreme cases, namely nonrelativistic ($\beta_j\ll1$) and ultra-relativistic ($\beta_j\gg1$)  fluid flows.  
\par It is believed that most compact astrophysical objects are immersed in a strong magnetic field. So, a possible extension of the present study could be to a magnetized relativistic multi-component degenerate plasma at finite temperature. The importance of the quantum effects like the particle dispersion and the particle spin may be examined and included, if necessary, in the extended model. These improvements, however, requires a further study, so that the model could fit with some laboratory experiments, to be designed, or some real astrophysical observations.
\section*{Author contributions statement}
R. D. and G. B. investigated, performed the numerical analysis, analyzed the results, and wrote the original draft (equal).  A. P. M. conceptualized, supervised, validated and analyzed the results, and edited the manuscript. C. B. investigated and contributed to analytical results. All authors reviewed and approved the manuscript. 
 %%%%%%%%%%%%%%%%%5
\section*{Data availability}
The datasets used and/or analysed during the current study are available from the corresponding author on reasonable request.

\section*{Acknowledgments}  One of us, RD, acknowledges support from the University Grants Commission (UGC), Government
of India, for a Junior Research Fellowship (JRF) with Ref. No. 1161/(CSIR-UGC NET DEC. 2018) and F. no. 16-6 (DEC. 2018)/2019 (NET/CSIR). This work was initiated and major parts were completed when Gadadhar Banerjee was on leave from the Department of Mathematics, University of Engineering \& Management (UEM), Kolkata-700 160, India, to work in the Department of Mathematics, Visva-Bharati University, India, under the Dr. D. S. Kothari Post Doctoral Fellowship Scheme of the University Grants Commission (UGC), Govt. of India with Ref. No. F.4-2/2006(BSR)/MA/18-19/0096).   

\bibliography{ref}

\end{document}